\documentclass[a4paper,12pt,twoside,english]{article}
\usepackage[latin1]{inputenc}
\usepackage{parskip}               
\usepackage[dvips]{graphicx,epsfig,color}
\usepackage{wrapfig,rotating}
\usepackage{amsmath,wasysym,array}
\usepackage[thinspace, thickqspace, amssymb, Gray]{SIunits}
\usepackage{hhline}
\usepackage{cite}                  
\usepackage[ , ,bf,it]{caption}    
 
\newlength{\dinwidth}
\newlength{\dinmargin}
\newcommand{\artitle}[1]{}
\newcommand{\artitlekeep}[1]{{\em``#1''}}

\newcommand{\whizard}      {{\sc Whizard}}
\newcommand{\whizomega}      {O'Mega}

\newcommand{\ra}            {\ensuremath{ \rightarrow     }}

\newcommand{\ep}{e^{+}e^{-}}
\newcommand{\ccg}{\chi\chi\gamma}
\newcommand{\nng}{\nu\bar{\nu}\gamma}

\newcommand{\lum}{\mathcal{L}}

\newcommand{\Pe}{\ensuremath{ P_{e^{-}}}}
\newcommand{\Pp}{\ensuremath{ P_{e^{+}}}}
\newcommand{\fb}{{\rm fb}^{-1}}

\newcommand{\equal}{{\bf "Equal"}}
\newcommand{\hel}{{\bf "Helicity"}}
\newcommand{\anti}{{\bf "Anti-SM"}}
\definecolor{dblack}   {rgb}{0.00, 0.00, 0.00}  
\definecolor{rred}     {rgb}{1.00, 0.00, 0.00}  

\begin{document}
\begin{titlepage}
  \begin{flushleft}
    {\tt DESY 12-117    \hfill    ISSN 0418-9833} \\
    {\tt June 2012}                  \\
  \end{flushleft}

  \vspace{1.0cm}
  \begin{center}
    \begin{Large}
      {\bfseries \boldmath Characterising WIMPs at a future $e^+e^-$ Linear Collider}

      \vspace{1.5cm}
      Christoph Bartels$^{1,2}$, Mikael Berggren$^1$, and Jenny List$^1$
    \end{Large}

    \vspace{.3cm}
    1- Deutsches Elektronen-Synchrotron DESY \\ 
    Notkestr. 85,  22607 Hamburg, Germany
    \vspace{.1cm} \\
    2- Universit\"at Hamburg, Institut f\"ur Experimentalphysik \\
    Luruper Chaussee 149,  22761 Hamburg, Germany
  \end{center}

  \vspace{1cm}

  \begin{abstract}
 
We investigate the prospects for detecting and measuring the parameters
of WIMP dark matter in a model independent way at the International
Linear Collider. The signal under study is direct WIMP pair production 
with associated initial state radiation $\ep \ra \ccg$.
The analysis accounts for the beam energy spectrum
of the ILC and the dominant machine induced backgrounds. The influence
of the detector parameters are incorporated by full simulation
and event reconstruction within the framework of the ILD detector concept.
We show that by using polarised beams, the detection potential is 
significantly increased by reduction of the dominant SM background
of radiative neutrino production $\ep \ra \nng$. The dominant sources
of systematic uncertainty are the precision of the polarisation 
measurement and the shape of the beam energy spectrum.
With an integrated luminosity of $\lum = 500\;\fb$ the helicity
structure of the interaction involved can be inferred, and
the masses and cross-sections can be measured with a relative accuracy 
of the order of $1\%$.

  \end{abstract}

  \vspace{1.0cm}
  \begin{center}
    To be submitted to EPJC
  \end{center}

\end{titlepage}

\section{Introduction}
 
Weakly interacting massive particles~(WIMPs) $\chi$ with masses in the order of 
$M_{\chi}\simeq 100$~GeV are among the favoured candidate particles to provide the observed 
cosmological abundance of dark matter. Complementary to direct and indirect detection 
experiments which look for signals of primordial WIMPs, colliders could 
produce WIMP particles under laboratory conditions. Since the WIMPs themselves leave 
collider experiments undetected due to their weak interaction with matter, collider 
searches typically rely on WIMPs appearing in cascade decays of  more heavy exotic 
particles, thus assuming a specific extension of the Standard Model of particle 
physics~(SM).

Direct WIMP production does not depend on the existence or kinematical accessibility 
of additional new particles. However it can be observed only when accompanied by 
wide-angle initial state radiation~(ISR). It has been shown~\cite{bib:birkedal} 
that for annihilation cross-sections compatible with the relic abundance 
of dark matter, the cross-section for radiative WIMP production at an electron-positron 
Linear Collider can be sufficiently large to observe this process above the irreducible
background from radiative neutrino production. The energy spectrum of the ISR photon
can be exploited to extract information on the WIMP mass and cross-section. The resulting 
discovery reach and mass resolutions have been studied before assuming either supersymmetric 
extension of the SM or the presence of universal extra-dimensions~\cite{bib:perelstein}.
In both cases the new partners of the electron can be exchanged in the $t$-channel and thus 
impact the shape of the photon energy spectrum. Based on four-vector smearing and taking into
account the radiative neutrino background and an overall systematic uncertainty of $0.3\%$, 
the two cases can be distinguished for a large range of the parameter space, and the mass of the 
electron partner can be measured far beyond the kinematic limit for its direct production. 

However, the shape of the photon spectra from signal and SM background are expected 
to be significantly distorted by various effects related to the detector, the reconstruction 
and the accelerator. The observation reach for the single photon signature at the International 
Linear Collider (ILC) has been investigated in full detector simulation in~\cite{bib:taikan}. 
Considering the statistical uncertainty only and assuming fully polarised beams, it has been 
shown that with an integrated luminosity of $500$~fb$^{-1}$ cross-sections down to about $12$~fb 
can be observed at the $5 \sigma$ level.

This publication studies the precision to which the WIMP mass and cross-section can be 
determined based on full detector simulation and taking into account all relevant backgrounds
as well as systematic uncertainties of the detector measurement and the limited knowledge of 
beam parameters. Instead of studying different explicit extensions of the SM, the generic 
parametrisation of the WIMP cross-section~\cite{bib:birkedal} is used here to investigate 
if the dominant partial wave of the WIMP production cross-section can be determined.

Beyond the WIMP approach, the results of this study are relevant for pair production of any invisible
or nearly invisible particle, whenever the ISR recoil method is applicable. 
 
This paper is organised as follows: In the next section, the model-indepedent 
Ansatz for radiative WIMP production as well as the signal charcteristics 
and the most important backgrounds are introduced.  Section~\ref{sec:expcond} discusses
the experimental conditions at the ILC and one of the proposed detector concepts, including
the resulting event selection criteria and systematic uncertainties. The achievable
precisions on the WIMP mass and its polarised and unpolarised cross-sections as well as
the prospects to determine the dominant partial wave of the WIMP production mechanism
are finally presented in Section~\ref{sec:results}.

\section{Radiative WIMP production in $e^+e^-$ collisions}

\begin{figure}[!h]
  \begin{minipage}[c]{0.4\linewidth}
    \caption{\label{Fig:WIMPPseudoFeyn} \it Illustration of the radiative WIMP pair production 
    mechanism. No explicit assumption on the WIMP-fermion interaction is made, thus only ISR off 
    the incoming particles is considered.}
  \end{minipage}\hspace*{2.5mm}
  \begin{minipage}[c]{0.6\linewidth}
  \setlength{\unitlength}{1.0cm}
\begin{picture}(4.0, 4.0)
    \put( 2.00, 0.40)  {\epsfig{file=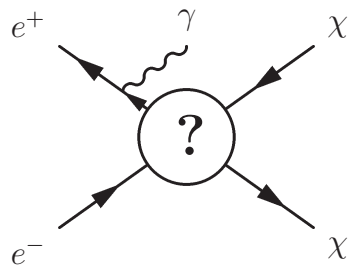, clip= , width=0.5\linewidth}}
\end{picture}
  \end{minipage}
\end{figure}

WIMPs can be pair produced in $e^+e^-$ collisions if they have a non-vanishing coupling
to electrons and if their mass doesn't exceed half of the center-of-mass energy. An additional
ISR photon allows to detect such events. In this analysis no specific scenario of physics beyond the SM is assumed and thus only radiation off the incoming particles is considered, as illustrated in  Figure~\ref{Fig:WIMPPseudoFeyn}. 

The production cross-section for WIMP pairs with an associated ISR photon of energy $E_{\gamma}$ and polar 
angle $\theta$  can be written in the limit of non-relativistic final state WIMPs as~\cite{bib:birkedal}:
\begin{equation}
\frac{d\sigma}{dx\; d\cos{\theta}}\approx \frac{\alpha\kappa_e\sigma_{\rm an}}{16\pi}
\frac{1+(1-x)^2}{x\sin^2{\theta}}2^{2J_0}(2S_{\chi}+1)^2 \left(1-\frac{4M_{\chi}^2}{(1-x)s}\right)^{1/2+J_0}
\label{Eqn:WIMP}
\end{equation}

Here, $M_{\chi}$ is the WIMP mass, $S_{\chi}$ its spin, $s$ the center-of-mass energy squared and the dimensionless 
variable $x=\frac{2E_{\gamma}}{\sqrt{s}}$. $J_{0}$ is the quantum number of the dominant partial wave
in the production process. The quantitiy $\kappa_{e}$ is the annihilation fraction of WIMPs into
electron positron pairs\footnote{With the sum over all final state fermions ($i = e, \nu, q,...$)
 $\sum_{i}\kappa_{i} = 1$.}. It implicitely depends on the helicity structure of the WIMP production mechanism
 and the helicities of the beam electrons and positrons. Under the additional assumption that the primordial dark matter consists of our WIMPs,  
the  overall scale of the production cross-section above is given by the WIMP annihilation cross-section into fermion-antifermion 
pairs $\sigma_{\rm an}$. It is estimated from the observed relic density to be about $0.8$~pb for s-wave and about $7$~pb for p-wave 
annihilation~\cite{bib:birkedal}.

\begin{figure}[htb]
  \begin{picture}(16.0, 8.0)
    \put( 0.00, 0.00)  {\epsfig{file=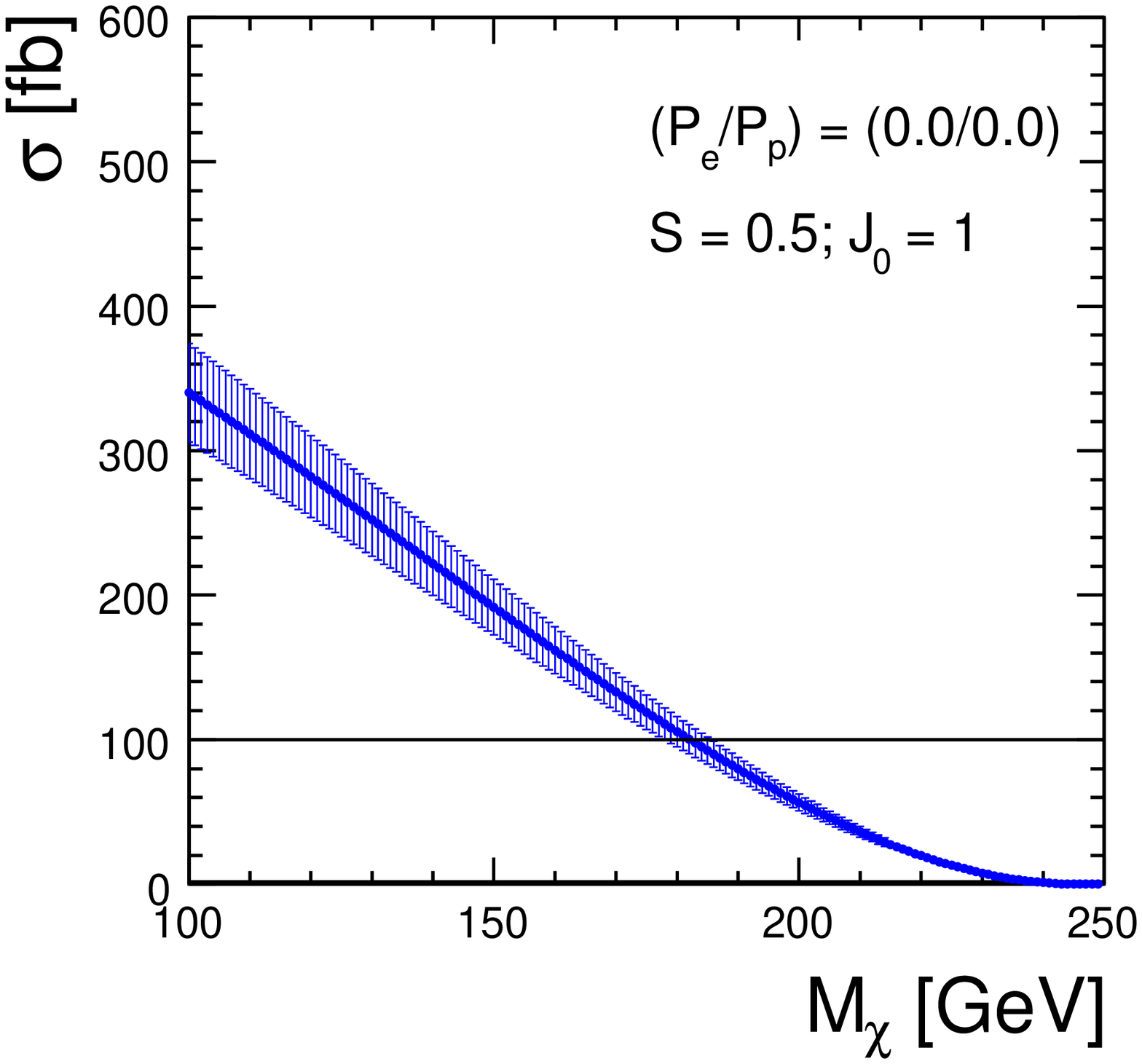, clip= , width=0.5\linewidth}}
    \put( 8.00, 0.00)  {\epsfig{file=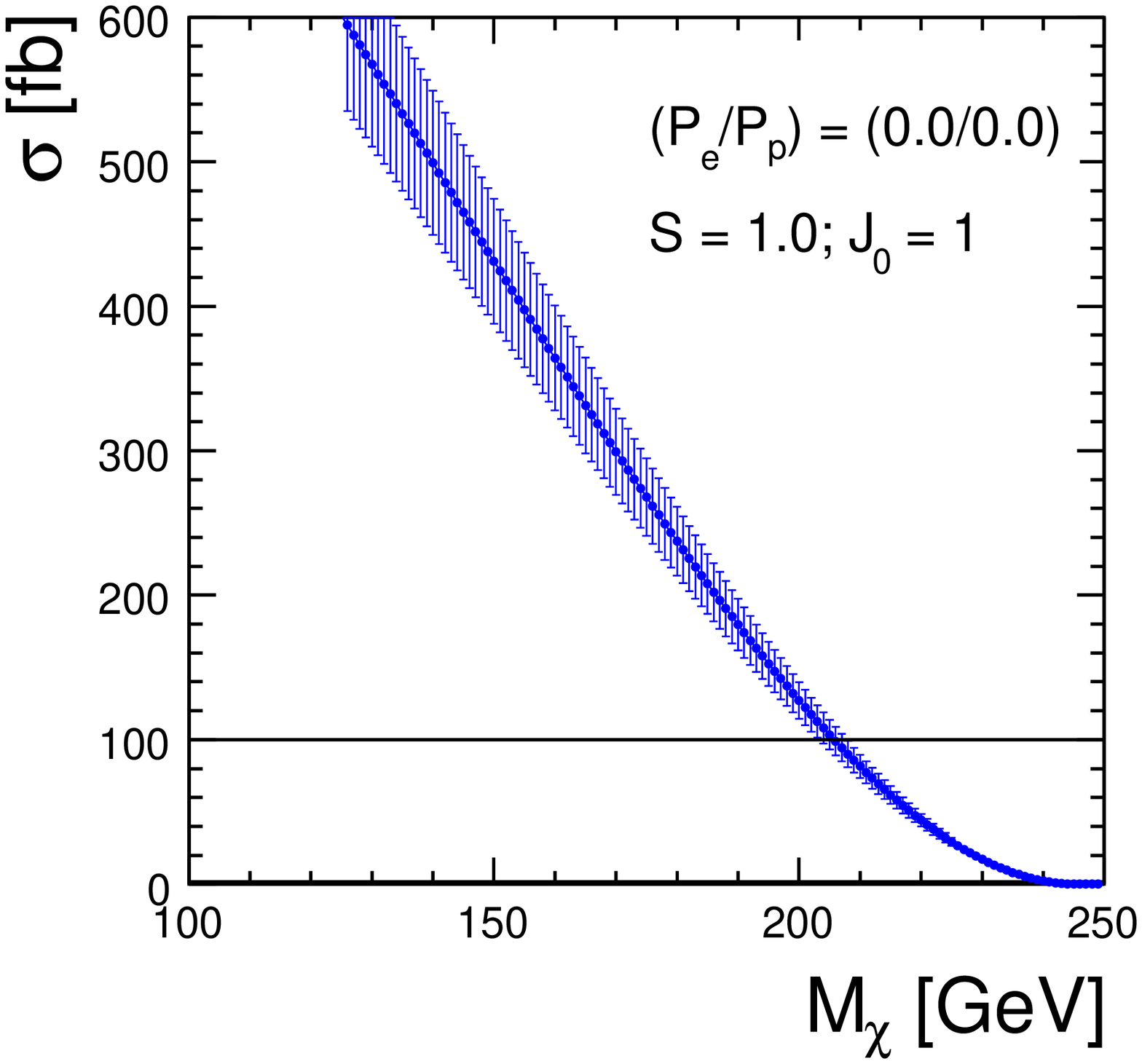, clip= , width=0.5\linewidth}}
    \put( 0.00, 0.00)  {(a)}
    \put( 8.00, 0.00)  {(b)}
  \end{picture}
  \caption{\label{fig:crosssection} \it Total visible cross-sections for radiative WIMP production at $500$~GeV for $\kappa_e = 1$: 
  (a) p-wave, spin-1/2, (b) p-wave, spin-1. The error bars illustrate an $10\%$ uncertainty.}
\end{figure}

Figure~\ref{fig:crosssection} shows the total unpolarised cross-section for $e^+e^-\rightarrow 
\chi \chi \gamma$ at a center-of-mass energy of $500$~GeV as a function of the WIMP mass for 
p-wave production and a (a) spin-1/2 or (b) spin-1  WIMP. The phase space of the photon has been 
restricted to an experimentally accessible range of $E_{\gamma} >8$~GeV and $|\cos{\theta}|<0.995$. 
The error bars illustrate a 10~\% uncertainty, the annihilation cross-sections have been taken from~\cite{bib:birkedal}. 
The cross-section is well above $100$~fb for WIMP masses up to $180$~GeV in the spin-1/2 and up to 
$200$~GeV in the spin-1 case. For $\kappa_{e}$ in the order of $10\%$ and large WIMP masses, the expected cross-sections 
get close to the minimum of $12$~fb for a $5~\sigma$ observation with an integrated luminosity of $500$~fb$^{-1}$ 
found in a previous study~\cite{bib:taikan}.

\begin{figure}[tb]
  \begin{picture}(16.0, 8.0)
    \put( 0.00, 0.00)  {\epsfig{file=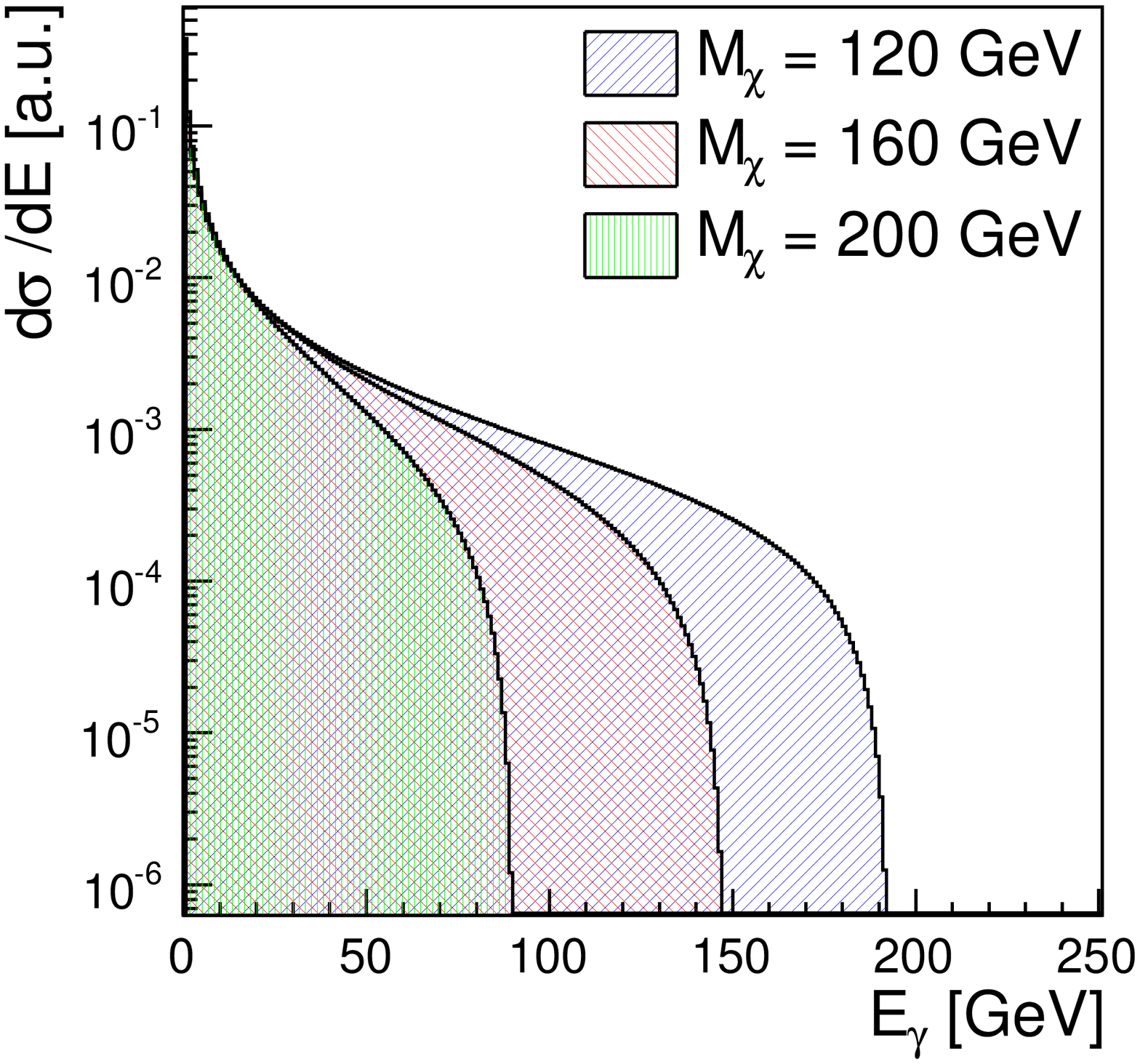, clip= , width=0.4\linewidth}}
    \put( 8.00, 0.00)  {\epsfig{file=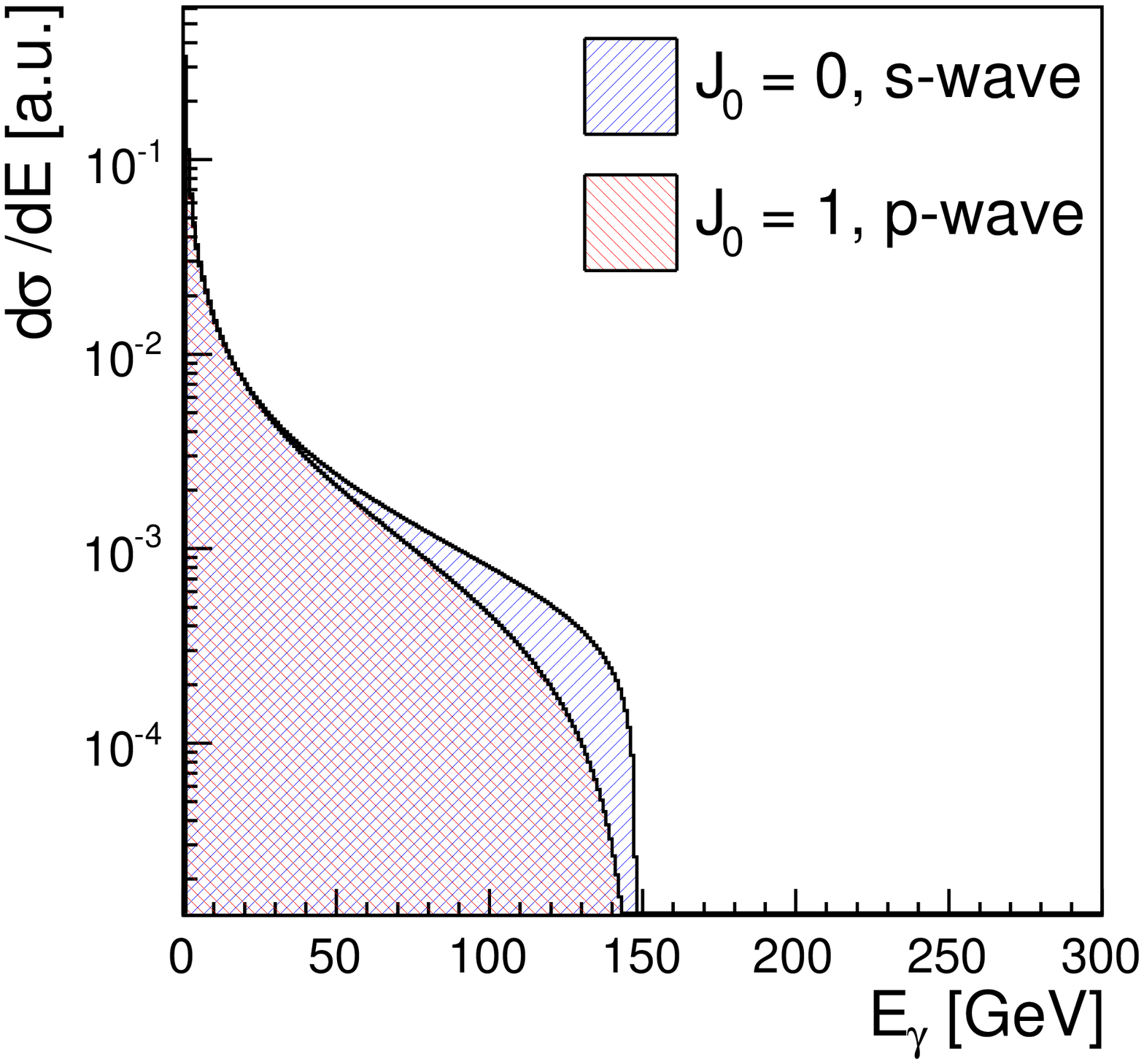, clip= , width=0.4\linewidth}}
    \put( 0.00, 0.00)  {(a)}
    \put( 8.00, 0.00)  {(b)}
  \end{picture}
  \caption{\label{fig:dsigmadE} \it Generator level photon energy spectra a) for different WIMP masses 
  in the p-wave case b) for $M_{\chi}= 160$~GeV in p- and s-wave production. All spectra are normalised
  to an integral of 1 to facilitate the shape comparison.}
\end{figure}

While the angular distribution of the ISR photons is independent of the following hard process, 
the photon energy spectrum strongly depends on the WIMP parameters. Figure~\ref{fig:dsigmadE}(a) 
shows the shape of the photon energy distributions expected for different WIMP masses in the case 
of p-wave production, featuring the mass dependent cut-off at the maximal 
allowed photon energy.  The shape of the photon energy threshold near its endpoint, which is
equivalent to the rise of the production cross-section at threshold, is compared in 
Figure~\ref{fig:dsigmadE}(b) for the  cases of $J_{0} =0$ (s-wave)  and $J_{0} =1$  (p-wave)  
production. Measurement of the threshold behaviour would provide an indication of the partial wave 
involved in the production process. This information could for instance be employed to
test whether the WIMP is a majorana fermion (like for instance a neutralino) or a scalar particle.

The single photon signature suffers from substantial irreducible background from radiative 
neutrino production $\ep\ra\nng$, which has an unpolarised cross-section in the order of several 
pb, depending on the acceptance cuts on the photon energy and angle. Due to the size of this 
background, a similar analysis at LEP would not have been feasible. At the ILC, the much higher 
luminosity and the beam polarisation help to gain sensitivity over the neutrino background, which 
proceeds primarily via t-channel $W$ exchange and hence is strongly polarisation dependent. Other 
important background processes comprise multi-photon final states $\ep\ra\gamma\gamma$ and radiative 
Bhabha scattering $\ep\ra\ep\gamma$,  as well as machine induced backgrounds.

\begin{table}[!h]
  \centering
  \renewcommand{\arraystretch}{1.10}
  \begin{tabular*}{\textwidth}{l@{\extracolsep{\fill}} r r r}
    \hline\hline 
     & \multicolumn{3}{c}{Cross-sections [fb] for $(\Pe;\Pp) = $} \\
    Process                           & $(-0.8;+0.3)$ &  $(+0.0;+0.0)$ & $(+0.8;-0.3)$ \\[0.5pt]
    \hline\hline 
     $\nng$                           &   $    5821$  &   $    2575$   &   $    1263$  \\
     $\nng\gamma$                     &   $   782.0$  &   $   355.4$   &   $     214$  \\
     $\nng\gamma\gamma$               &   $    55.8$  &   $    26.2$   &   $      19$  \\ \hline
     $\gamma\gamma$                   &   \multicolumn{3}{c}{$ 11.4 \times 10^{3}  $ }\\ 
     $\gamma\gamma\gamma$             &   \multicolumn{3}{c}{$ 1.1 \times 10^{3}  $ }\\ 
     $\gamma\gamma\gamma\gamma$       &   \multicolumn{3}{c}{$ 0.1 \times 10^{3}  $ }\\ \hline
     $e^+e^- $                        &   \multicolumn{3}{c}{$ 890 \times 10^{3}  $ }\\
    \hline \hline
  \end{tabular*}
  \caption{\it Cross-sections of the most important Standard Model backgrounds for three 
  different polarisation configurations. All photons denoted explicitely in the process name 
  are included in the matrix element. In addition two ISR photons are included. At least 
  one photon is required to have $E_{\gamma}>8$~GeV and $|\cos{\theta}|<0.995$.}
  \label{tab:xsections}
\end{table}

Table~\ref{tab:xsections} shows the cross-sections of these Standard Model backgrounds 
for three different beam polarisation configurations as obtained from \whizard~\cite{bib:whizard}.
\Pe and \Pp denote the values of the electron and positron beam polarisation, respectively. 
All photons denoted explicitely in the process name are included in the matrix element calculation
performed by \whizomega~\cite{bib:omega}. At least one photon is required to be in the 
analysis phase space as defined above. In each case two additional soft ISR photons are 
included in the event generation.

The comparison of these numbers to the expected signal cross-sections displayed in 
Figure~\ref{fig:crosssection} immediately demonstrates the need for a careful evaluation of the
experimental possibilities to suppress all reducible background processes. Furthermore the  
systematic uncertainties associated with the remaining irreducible background need to be understood
in order to obtain a realistic picture of the ILC's capabilities in this channel.

Since on an event-by-event basis the radiative WIMP production is indistinguishable from radiative
neutrino production, the signal sample was generated by reweighting one third of the 
$\nng$ sample according to the ratio of the tree-level differential cross-sections of WIMP 
and neutrino production. This method has the benefit of giving flexible access to the all 
combinations of WIMP properties without the need for dedicated Monte-Carlo production at each 
considered point in parameter space.

\section{Experimental Conditions}
\label{sec:expcond}

This study is based on an integrated luminosity of $500$~fb$^{-1}$ and the nominal parameter set 
of the ILC at a center-of-mass energy of $500$~GeV as specified in its Reference Design 
Report~\cite{bib:RDR}. For comparison, the SB2009 beam parameter set~\cite{bib:SB2009} 
is used for selected results. The chosen beam parameter sets differ mainly in the strength of 
the focussing of the beam at the interaction point, resulting in different beam energy spectra, 
which have been derived using GuineaPig~\cite{bib:guineapig}. 

The electron beam polarisation of $|P_{e^-}| = 80\%$ and the positron polarisation of 
$|P_{e^+}| = 30\%$ are assumed to be known to $\delta P /P = 0.25\%$~\cite{bib:EP-paper}. 
In order to estimate the impact of these quantities, alternative values of  $|P_{e^+}| = 60\%$ 
and $\delta P /P = 0.1\%$ are used for comparison. The peak beam energy will be measured to a 
relative precision of $10^{-4}$~\cite{bib:EP-paper} by the energy spectrometers. For the evaluation
of the systematic uncertainties in this analysis, a slightly more conservative estimate based on
the position of the radiative return to the $Z$ in the photon energy spectrum is used. Assuming the
same amount of integrated luminosity as used in the main analysis, this position can be determined
to a statistical precision of $80$~MeV. For the integrated luminosity a relative precision of 
$10^{-3}$ is assumed~\cite{bib:lumi}.

The analysis is performed in full simulation of the ILD detector concept~\cite{bib:ILDLoI}. 
The central device for detecting the ISR photons is the electromagnetic calormeter (ECAL), 
which is a highly granular SiW sampling calorimeter with a cell size of $5\times5$~mm$^2$.
In testbeam measurements~\cite{bib:ecalperf}, a resolution of  

\begin{equation}
 \frac{\Delta E}{E} = (16.6 \pm 0.1)\%\times \frac{1}{\sqrt{E}} \oplus (1.1\pm 0.1)\%,
\label{Eq:Eres}
\end{equation}

as has been achieved. The simulation used here shows a similar performance.

\begin{figure}[htb]
  \begin{picture}(16.0, 8.0)
    \put( 0.00, 0.00)  {\epsfig{file=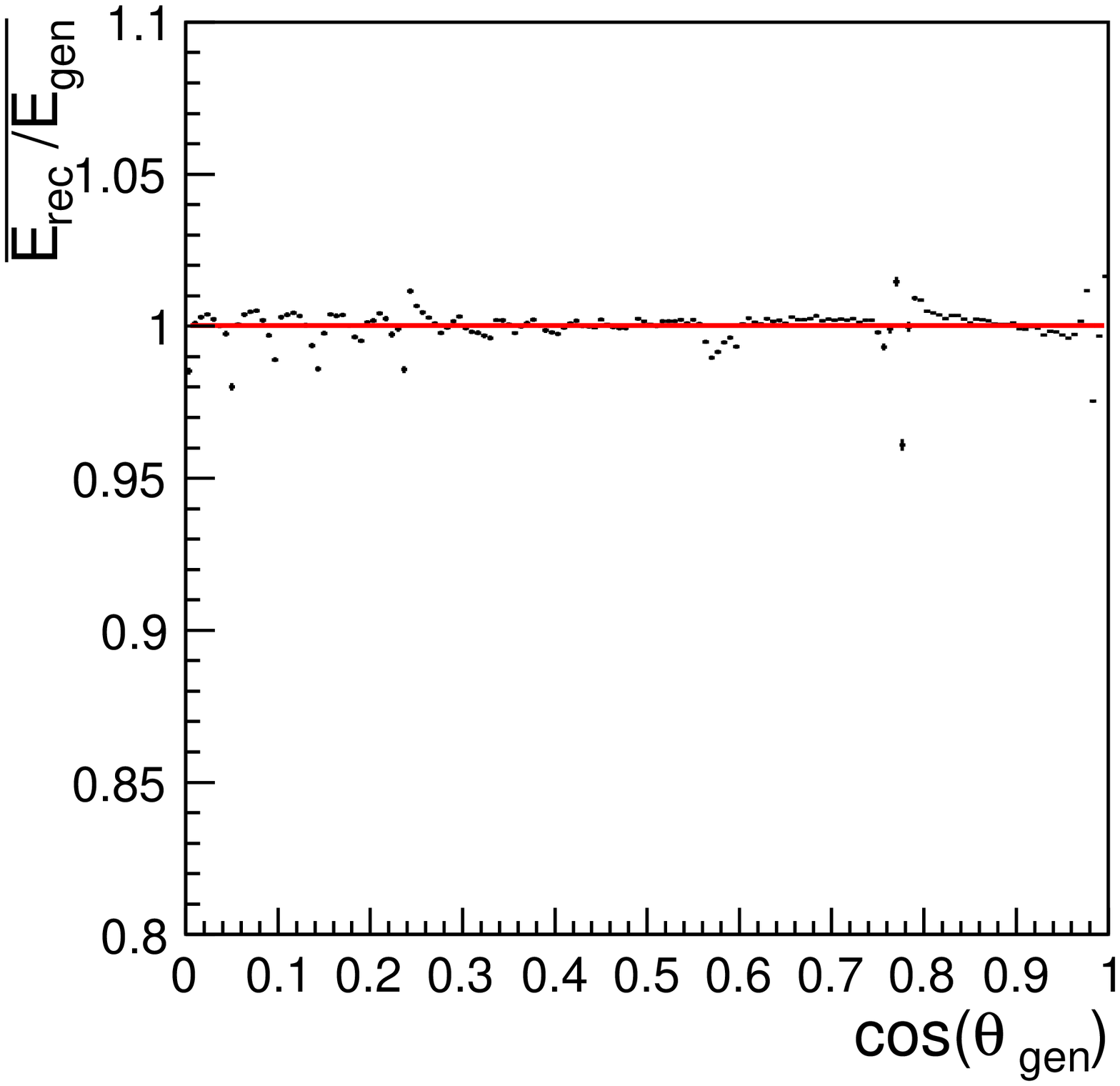, clip= , width=0.4\linewidth}}
    \put( 8.00, 0.00)  {\epsfig{file=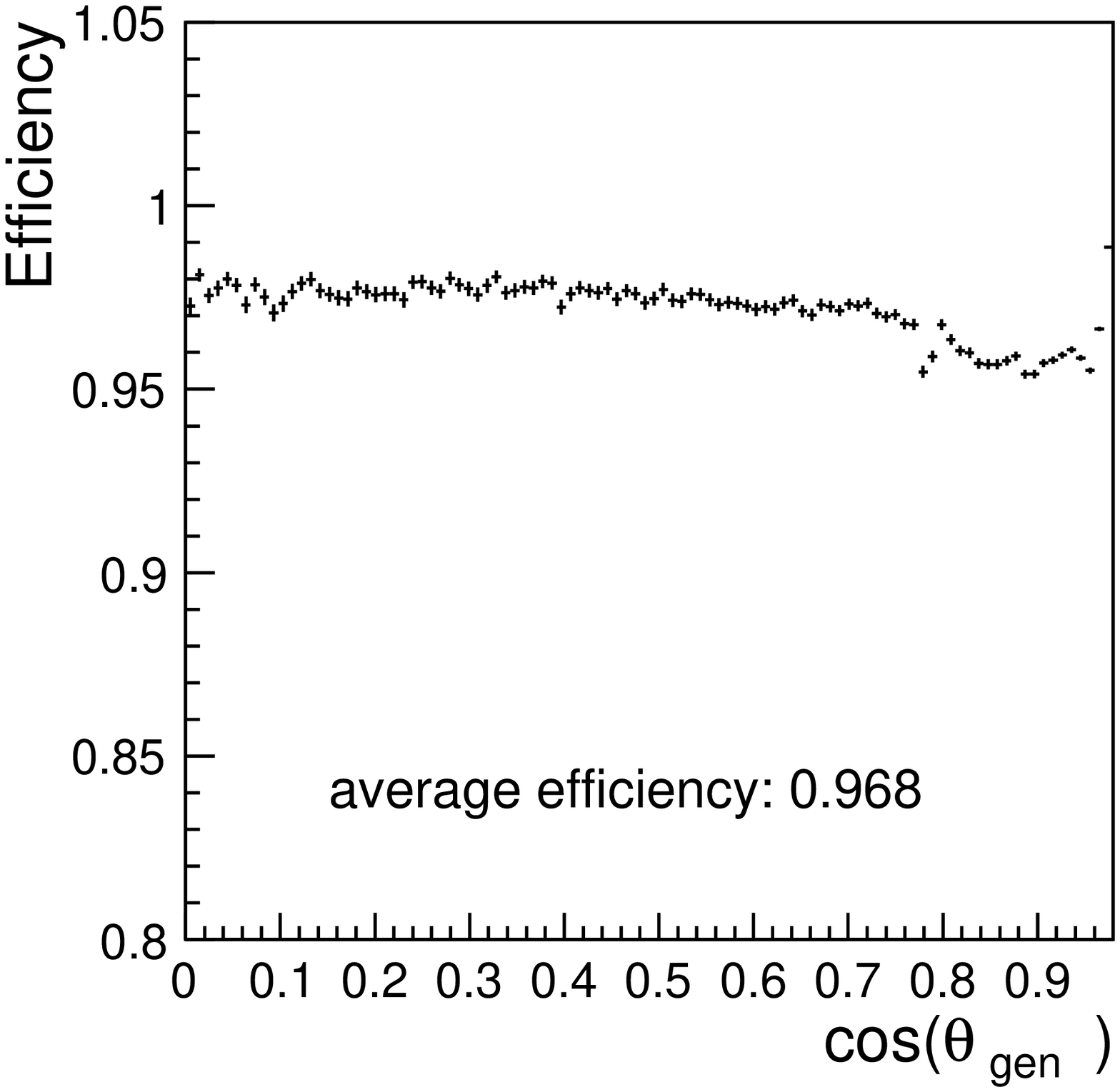, clip= , width=0.4\linewidth}}
    \put( 0.00, 0.00)  {(a)}
    \put( 8.00, 0.00)  {(b)}
  \end{picture}
  \caption{\label{fig:efficiency} \it (a) average of reconstructed over true photon energy as
   function of $\cos{\theta}$ after corrections for cluster splitting and losses in 
   insensitive regions of the detector. \newline
   (b) reconstruction efficiency for isolated photons as 
   function of $\cos{\theta}$.}
\end{figure}

ECAL clusters to which no track can be associated are considered as photon candidates. After 
corrections for split clusters and losses in the cracks between modules, the energy of isolated
photons can be reconstructed without bias over the whole polar angle range. This is illustrated by 
Figure~\ref{fig:efficiency}(a), which shows the ratio of reconstructed over true photon energy as 
function of $\cos{\theta}$ of the photon. Residual deviations from unity near module boundaries are 
smaller than $2\%$. Figure~\ref{fig:efficiency}(b) shows the efficiency of the photon reconstruction 
as a function of $\cos{\theta}$. The average efficiency is near $97\%$, varying from close to $98\%$ 
in the middle of the detector to about $95\%$ in the endcaps. For further analysis, only events with 
at least one photon with a reconstructed energy $10$~GeV $< E_{\gamma}^{\mathrm{rec}} < 220$~GeV  and
a polar angle $|\cos{\theta^{\mathrm{rec}}}| < 0.98$ are considered. 

In order to distinguish photons from electrons or hadrons and to veto any other significant detector 
activity, the tracking system and the scintillator-tile steel sandwich hadronic calorimeter (AHCAL) are 
employed in addition. Both ECAL and AHCAL as well as the tracking system are placed inside a 
superconducting coil which provides a $3.5$~T solenoidal magnetic field. The tracking system combines 
a Time Projection Chamber~(TPC) with additional silicon pixel and strip detectors to achieve a 
tracking efficiency of $\ge 99.5\%$ in an angular range of $7^{\circ} < \theta < 173^{\circ} \quad 
(|\cos{\theta}| < 0.9925)$. The veto on tracks of charged particles needs to be adjusted to allow for 
possible tracks from beam background or low $p_t$ photon-photon interactions. 
\begin{figure}[!h]
  \begin{minipage}[c]{0.4\linewidth}
    \caption {\label{fig:ptbackground} \it Transverse momentum distribution of tracks from 
    $e^+e^-$ pair 
    background and from $\gamma\gamma$ interactions. The distributions are normalised to the 
    number of tracks expected per bunch crossing from these sources.}
  \end{minipage}\hspace*{2.5mm}
  \begin{minipage}[c]{0.6\linewidth}
  \setlength{\unitlength}{1.0cm}
  \begin{picture}(8.0, 7.0)
    \centerline {\epsfig{file=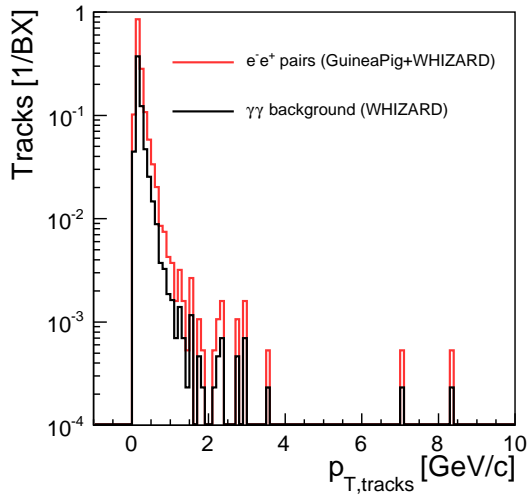, clip= , width=0.75\linewidth}}
  \end{picture}
  \end{minipage}
\end{figure}

The transverse momentum distribution of these tracks obtained from full detector simulation and 
reconstruction is displayed in Figure~\ref{fig:ptbackground}, 
together with the corresponding distribution for tracks originating from $e^+e^-$ pair background. 
Both distributions are normalised to the number of tracks expected from these processes per bunch 
crossing (BX), i.e. $0.7$ tracks/BX from $\gamma\gamma$ processes and $1.5$ tracks/BX from $e^+e^-$ pairs.
In the analysis, events containing any track with $p_T > 3$~GeV are vetoed. The remaining probability 
to falsely reject a signal event due to an overlayed background track is about $0.25\%$. In addition
to the track veto, events with other neutral particles are rejected by requiring that the total 
visible energy of the event must not exceed the recontructed photon energy by more than $20$~GeV.

The low angle calorimeters complement the hermeticity of the detector, leaving only holes of 
$4.5$~mrad and $5.6$~mrad in the polar angle coverage of the BeamCal around the incoming and outgoing beams, respectively. 
The innermost part of the BeamCal is exposed to energy depositions of several TeV per bunch crossing 
from $e^+e^-$ pairs created by Beamstrahlung processes. Isolated high energy electrons (or photons) can 
be detected above this background with efficiencies larger than $80\%$ for angles above typically 
$20$~mrad from the detector $z$ axis\footnote{The detector $z$ axis points in the middle between 
incoming and outgoing beams, which cross under an angle of $14$~mrad.}~\cite{bib:beamcal}. The 
complete angular and energy dependence of this efficiency has been taken into account in the 
simulation. This is crucial in order to obtain a correct estimate of the background from radiative 
Bhabha scattering, since vetoing such isolated high energy electrons in the BeamCal is mandatory to 
reduce the Bhabha background to a manageable level.

\begin{table}[!h]
  \centering
  \renewcommand{\arraystretch}{1.10}
  \begin{tabular*}{\textwidth}{l@{\extracolsep{\fill}}  r rrr r}
    \hline\hline 
    \multicolumn{6}{c}{\quad } \\[-4.9mm]
    Process     \quad  \quad  &   $E_{\gamma} ; \cos{\theta}$ & $p_{T,track}$ & $E_{vis}-E_{\gamma}$ & BeamCal tag & Eff. [\%]\\[0.5pt]
    \hline\hline 
   $\nng$                     &   $  2493.3$  &   $ 2435.4$  &   $ 2283.88$  &  $2239.63$ & $89.8$\\
   $\nng\gamma$               &   $   344.3$  &   $  325.4$  &   $  238.52$  &  $ 228.51$ & $66.4$\\
   $\nng\gamma\gamma$         &   $    25.4$  &   $   23.2$  &   $   11.82$  &  $  11.05$ & $43.5$\\
   $\gamma\gamma$             &   $   578.1$  &   $  457.3$  &   $   60.74$  &  $   5.80$ & $ 1.0$\\
   $\gamma\gamma\gamma$       &   $   145.0$  &   $  112.7$  &   $    4.65$  &  $   0.10$ & $ 0.1$\\
   $\gamma\gamma\gamma\gamma$ &   $    19.5$  &   $   14.7$  &   $    0.15$  &  $   0.03$ & $ 0.2$\\
   $e^+e^- $                  &   $421533.1$  &   $88935.9$  &   $67389.80$  &  $1228.70$ & $ 0.3$\\
    \hline \hline
  \end{tabular*}
  \caption{\it Number of events for the main background processes at the different selection stages 
    for an integrated luminosity of $\lum = 1\,\fb$ and $(\Pe;\Pp) = (+0.0;+0.0)$. 
    The second column contains the event numbers in the analysis phace space. 
    The third column lists the event numbers after the cuts on the energy and polar angle of the
    reconstructed photon. The last column lists the selection efficiencies.}
  \label{tab:cutflow}
\end{table}

Table~\ref{tab:cutflow} illustrates the effect of the selection cuts for the case of unpolarised beams.
Beam polarisation significantly reduces or enhances the rates for the neutrino processes while leaving
the Bhabha rate at the same level. The final selection efficiency for the missing energy plus photon
signature is near $90\%$. In real data, this efficiency could be controlled from the high energy
part of the photon energy spectrum which contains the radiative return to the $Z$ and is not used in 
the WIMP analysis as well as from radiative muon pair production $\mu^+\mu^-\gamma$. From the 
statistics available in these two processes, we estimate the systematic uncertainty on the selection
efficiency to $1.5\%$. In addition, the impact of the cut on the reconstructed photon energy depends
slightly on the WIMP mass. Based on the accuracies on the WIMP mass achieved in this study, this
translates into an additional sytematic effect on the selection efficiency. Added in quadrature, both
effects together yield a total systematic uncertainty of $1.75\%$ on the selection efficiency.

\begin{figure}[htb]
  \begin{picture}(16.0, 8.0)
    \put( 0.00, 0.00)  {\epsfig{file=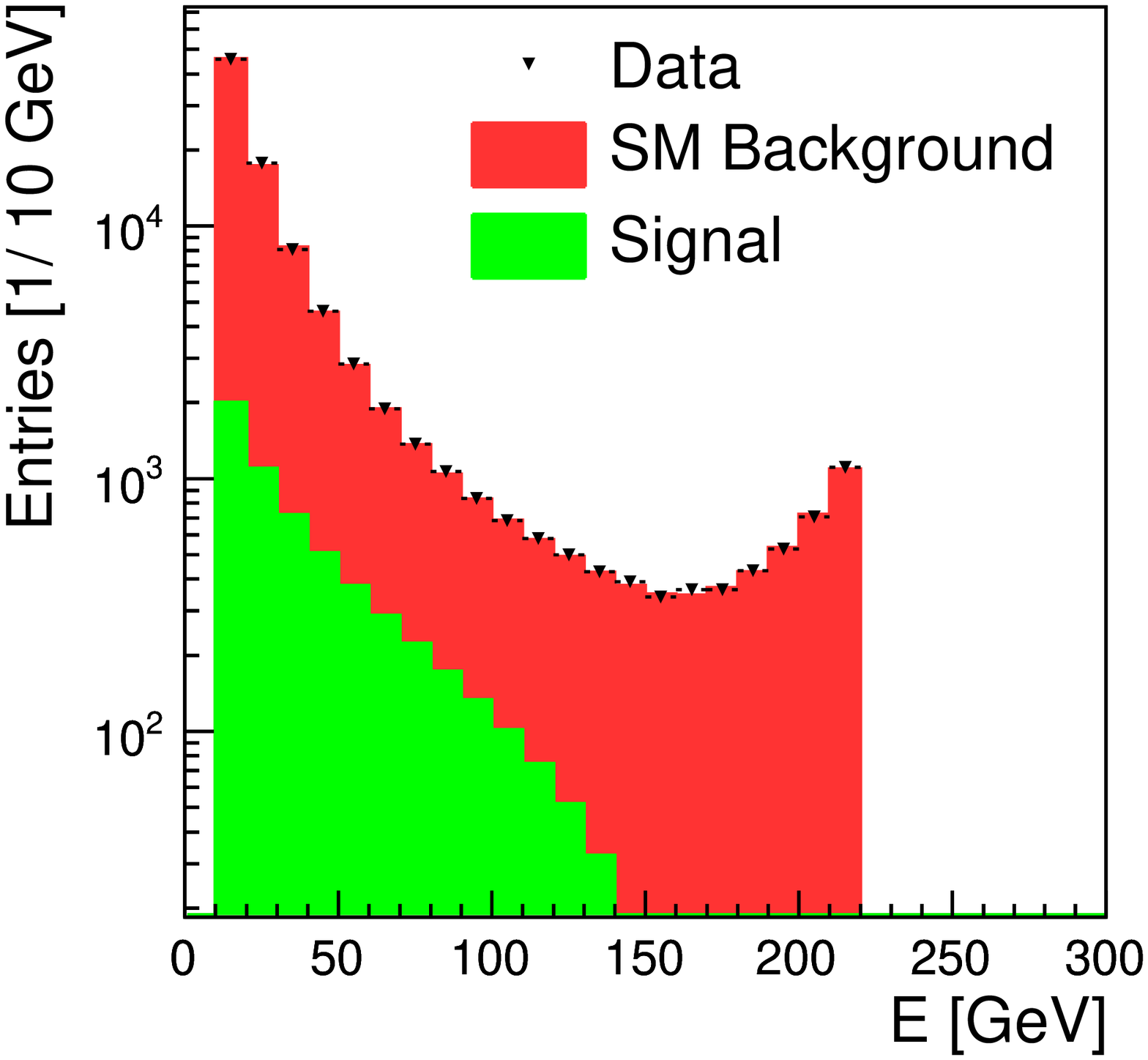, clip= , width=0.5\linewidth}}
    \put( 8.00, 0.00)  {\epsfig{file=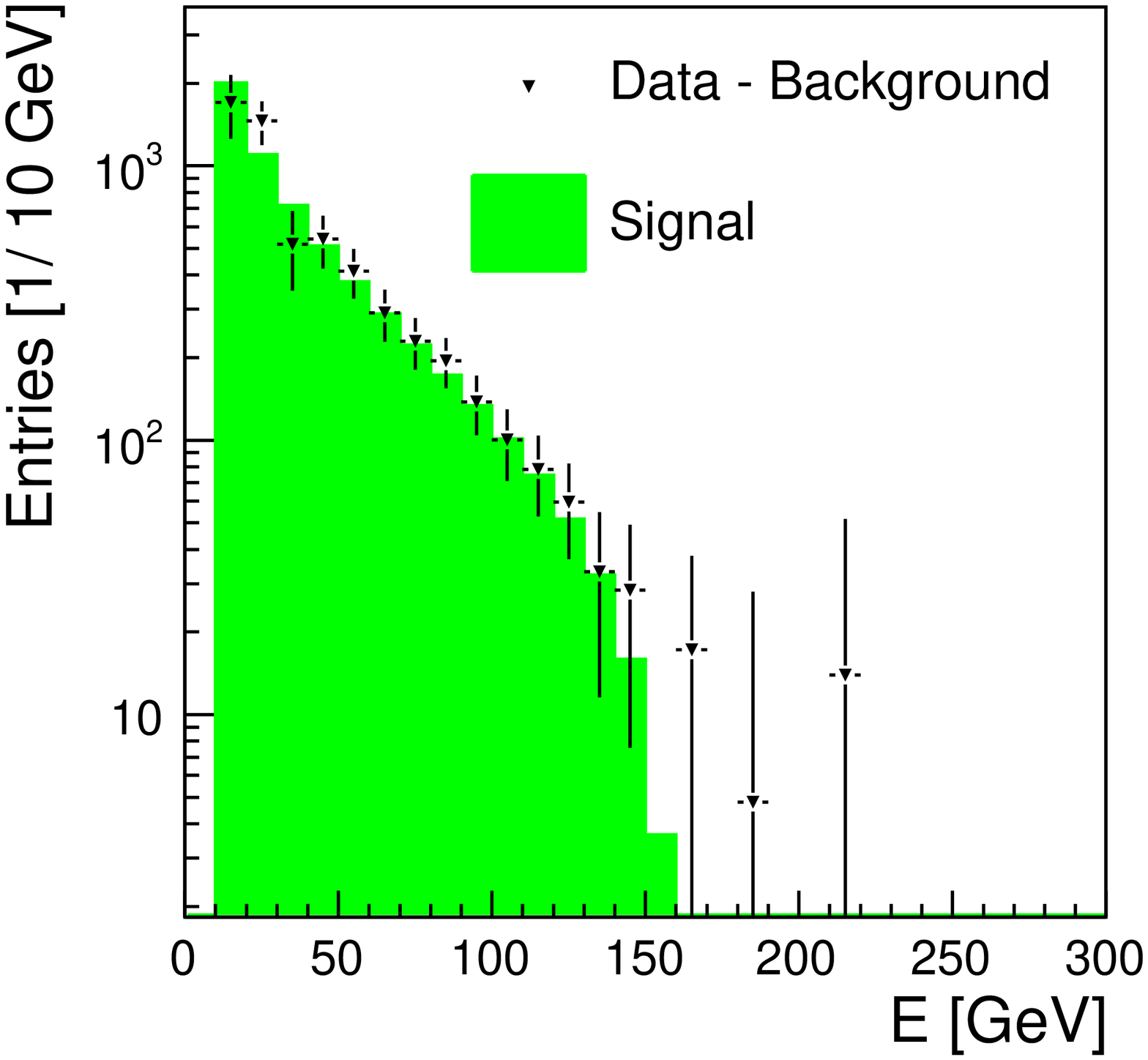, clip= , width=0.5\linewidth}}
    \put( 0.00,0.00)  {(a)}
    \put( 8.00,0.00)  {(b)}
  \end{picture}
  \caption{\label{fig:egammaSB} \it Photon energy spectra after all selection cuts for Standard Model
  background and a $150$~GeV p-wave WIMP signal, normalised to an integrated luminosity of $50$~fb$^{-1}$ and beam 
polarisations of $(\Pe;\Pp) = (+0.8;-0.3)$. (a) Background (red), signal (green) and a statistically
independent ``data'' sample (points) (b) Signal and ``data'' after background subtraction.}
\end{figure}

The photon energy spectrum obtained after all selection cuts is displayed in 
Figure~\ref{fig:egammaSB}(a) for Standard Model background (red) and for a WIMP 
signal (green; $M_{\chi} = 150$~GeV, p-wave). The distributions are normalised to 
an integrated luminosity of $50$~fb$^{-1}$ and beam polarisations of $(\Pe;\Pp) = (+0.8;-0.3)$. 
The ``data'' points are obtained from a statistically independent Monte-Carlo sample assuming 
the presence of background and signal. Figure~\ref{fig:egammaSB}(b) shows the ``data'' after 
subtracting the expected background compared to the signal expectation (green). The size of the 
statistical errors of the ``data'' points reflects the 
fluctuations of the large amount of subtracted background.

\section{Results}
\label{sec:results}

In this section the precisions on the polarised and unpolarised cross-sections and on the WIMP
mass achievable at the ILC are presented. Additionally the possibility to discriminate between
s- and p-wave production is investigated. In the model-independent approach, the branching fraction
of WIMP annihilation into $e^+e^-$ pairs, $\kappa_{e}$, has a free dependency on the helicity of 
the initial state electrons. In order to illustrate the power of polarised beams to determine this
dependency, three different coupling scenarios are compared in the following:
\begin{itemize}
\item \equal: The WIMP couplings are independent of the
      helicity of the incoming electrons and positrons, 
      i.e.~$\kappa(e^{-}_R,e^{+}_L) = \kappa(e^{-}_R,e^{+}_R) = \kappa(e^{-}_L,e^{+}_L) = \kappa(e^{-}_L,e^{+}_R)$.
\item \hel: The couplings conserve helicity and parity,\\
      $\kappa(e^{-}_R,e^{+}_L)  = \kappa(e^{-}_L,e^{+}_R)$; $\;\;\;\kappa(e^{-}_R,e^{+}_R) = \kappa(e^{-}_L,e^{+}_L) = 0$.
\item \anti: This scenario is a "best case" scenario, since
      the WIMPs couple only to right-handed electrons and left-handed positrons:
      $\kappa(e^{-}_R,e^{+}_L)$, with all other \mbox{$\kappa(e^{-},e^{+}) = 0$}.
\end{itemize}

\subsection{Cross-section Measurements}

For this first part of the analysis a typical running scenario of the ILC is assumed,
where an integrated luminosity of $\lum = 500\;\fb$ is split into $200\;\fb$ each with 
$(+|\Pe|;-|\Pp|)$ and $(-|\Pe|;+|\Pp|)$ (short $(+-)$ and $(-+)$), as well as into 
$50\;\fb$ each with  $(+|\Pe|;+|\Pp|)$ and $(-|\Pe|;-|\Pp|)$ (short $(++)$ and $(--)$).

By combining the four independent measurements with different beam
polarisation configurations, the helicity structure of the WIMP
interactions can be extracted by solving the equation system 

\begin{eqnarray}\label{Eqn:System}
 \sigma_{+-} & = &\frac{1}{4}\lbrace (1+|\Pe|)(1-|\Pp|) \sigma_{RR} + (1-|\Pe|)(1+|\Pp|) \sigma_{LL}\nonumber\\
                        && + \,\,\, (1+|\Pe|)(1+|\Pp|) \sigma_{RL} + (1-|\Pe|)(1-|\Pp|) \sigma_{LR} \rbrace\nonumber\\
 \sigma_{-+} & = &\frac{1}{4}\lbrace (1-|\Pe|)(1+|\Pp|) \sigma_{RR} + (1+|\Pe|)(1-|\Pp|) \sigma_{LL}\nonumber\\
                        && + \,\,\, (1-|\Pe|)(1-|\Pp|) \sigma_{RL} + (1+|\Pe|)(1+|\Pp|) \sigma_{LR} \rbrace\nonumber\\
 \sigma_{++} & = &\frac{1}{4}\lbrace (1+|\Pe|)(1+|\Pp|) \sigma_{RR} + (1-|\Pe|)(1-|\Pp|) \sigma_{LL}\nonumber\\
                        && + \,\,\, (1+|\Pe|)(1-|\Pp|) \sigma_{RL} + (1-|\Pe|)(1+|\Pp|) \sigma_{LR} \rbrace\nonumber\\
 \sigma_{--} & = &\frac{1}{4}\lbrace (1-|\Pe|)(1-|\Pp|) \sigma_{RR} + (1+|\Pe|)(1+|\Pp|) \sigma_{LL}\nonumber\\
                        && + \,\,\, (1-|\Pe|)(1+|\Pp|) \sigma_{RL} + (1+|\Pe|)(1-|\Pp|) \sigma_{LR}\rbrace,
\end{eqnarray}

In practice, $|\Pe|$ (as well as $|\Pp|$) will not be exactly equal for the four running periods, so that 
the measured cross-sections need to be extrapolated to the same $|\Pe|$ ($|\Pp|$) based on the actual 
polarimeter measurements and their systematic uncertainty. 

\begin{table}[!h]
  \centering
  \renewcommand{\arraystretch}{1.08}
  \begin{tabular*}{\textwidth}{l@{\extracolsep{\fill}} p{2mm} rr}  
    \hline\hline 
    \multicolumn{4}{c}{\quad } \\[-4.9mm]
    \quad                    && $(|\Pe|;|\Pp|) = (0.8;0.3)$ & $(|\Pe|;|\Pp|) = (0.8;0.6)$ \\[0.5pt]
    \hline\hline 
    \multicolumn{4}{c}{\quad} \\[-2mm]
    \multicolumn{4}{l}{\quad{\bf"Equal"}\quad scenario } \\ \hline
    \multicolumn{4}{c}{\quad} \\[-4.9mm]
    $\sigma_{RL}/\sigma_{0}$ &&    $0.99 \pm 0.24$ \quad $(0.16)$  &   $ 0.99 \pm 0.10$ \quad $(0.07)$ \\
    $\sigma_{RR}/\sigma_{0}$ &&    $1.00 \pm 0.33$ \quad $(0.21)$  &   $ 1.00 \pm 0.23$ \quad $(0.14)$ \\
    $\sigma_{LL}/\sigma_{0}$ &&    $1.00 \pm 0.37$ \quad $(0.29)$  &   $ 1.00 \pm 0.23$ \quad $(0.15)$ \\
    $\sigma_{LR}/\sigma_{0}$ &&    $0.95 \pm 0.38$ \quad $(0.25)$  &   $ 0.95 \pm 0.28$ \quad $(0.15)$ \\
    \hline 
    \multicolumn{4}{c}{\quad} \\[-2mm]
    \multicolumn{4}{l}{\quad{\bf\hel}\quad scenario } \\ \hline
    \multicolumn{4}{c}{\quad} \\[-4.9mm]
    $\sigma_{RL}/\sigma_{0}$ &&    $1.99 \pm 0.24$ \quad $(0.16)$  &   $ 1.99 \pm 0.10$ \quad $(0.08)$ \\
    $\sigma_{RR}/\sigma_{0}$ &&    $0.00 \pm 0.33$ \quad $(0.21)$  &   $ 0.00 \pm 0.23$ \quad $(0.14)$ \\
    $\sigma_{LL}/\sigma_{0}$ &&    $0.00 \pm 0.37$ \quad $(0.29)$  &   $ 0.00 \pm 0.23$ \quad $(0.15)$ \\
    $\sigma_{LR}/\sigma_{0}$ &&    $1.95 \pm 0.38$ \quad $(0.25)$  &   $ 1.95 \pm 0.29$ \quad $(0.16)$ \\
    \hline 
    \multicolumn{4}{c}{\quad} \\[-2mm]
    \multicolumn{4}{l}{\quad{\bf\anti}\quad scenario } \\ \hline
    \multicolumn{4}{c}{\quad} \\[-4.9mm]
    $\sigma_{RL}/\sigma_{0}$ &&    $3.99 \pm 0.26$ \quad $(0.18)$  &   $ 3.99 \pm 0.12$ \quad $(0.10)$ \\
    $\sigma_{RR}/\sigma_{0}$ &&    $0.00 \pm 0.33$ \quad $(0.22)$  &   $ 0.00 \pm 0.23$ \quad $(0.14)$ \\
    $\sigma_{LL}/\sigma_{0}$ &&    $0.00 \pm 0.36$ \quad $(0.28)$  &   $ 0.00 \pm 0.23$ \quad $(0.15)$ \\
    $\sigma_{LR}/\sigma_{0}$ &&   $-0.05 \pm 0.37$ \quad $(0.24)$  &   $-0.05 \pm 0.28$ \quad $(0.15)$ \\
    \hline 
  \end{tabular*}
  \caption[Measurement results of fully polarised cross-sections] {\it 
    Fully polarised cross-sections $\sigma_{\lbrace R,L\rbrace}$ measured within 
    three WIMP scenarios and for two different absolute polarisations of electrons and positrons. 
    The values are normalised to the input unpolarised cross-section of $\sigma_0=100\;$fb.
    The quoted uncertainties are the squared sum of statistical errors and systematic uncertainties, 
    with the bracketed values corresponding to an increased precision 
    on the polarisation measurement of $\delta P/P = 0.1\%$.
  }
  \label{Table:XSecPolMeas1}
\end{table}

\begin{figure}[hp]
  \begin{picture}(16.0, 21.0)
  \put(0.0,14.0){\epsfig{file=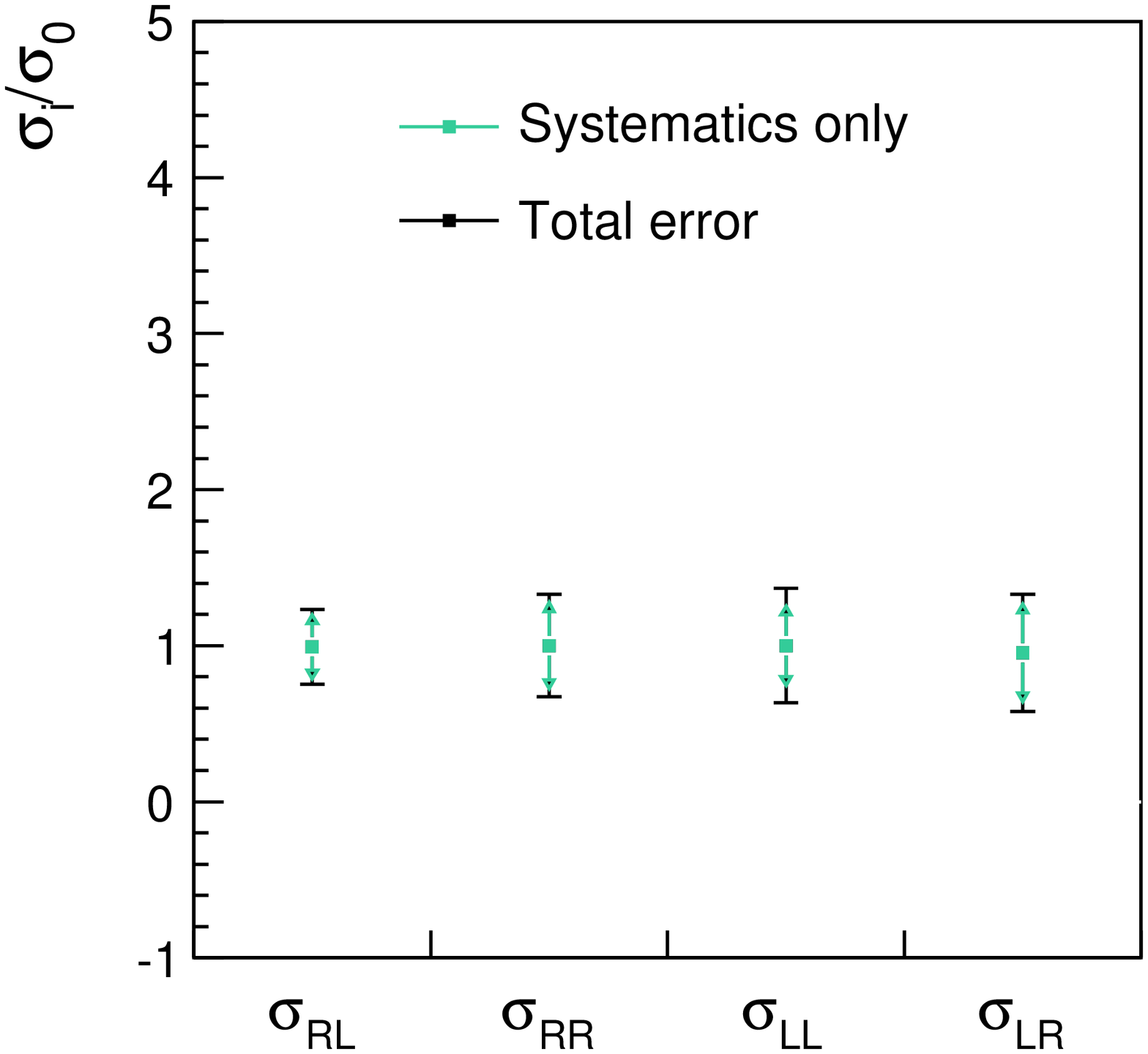,scale=.38}}
  \put(8.5,14.0){\epsfig{file=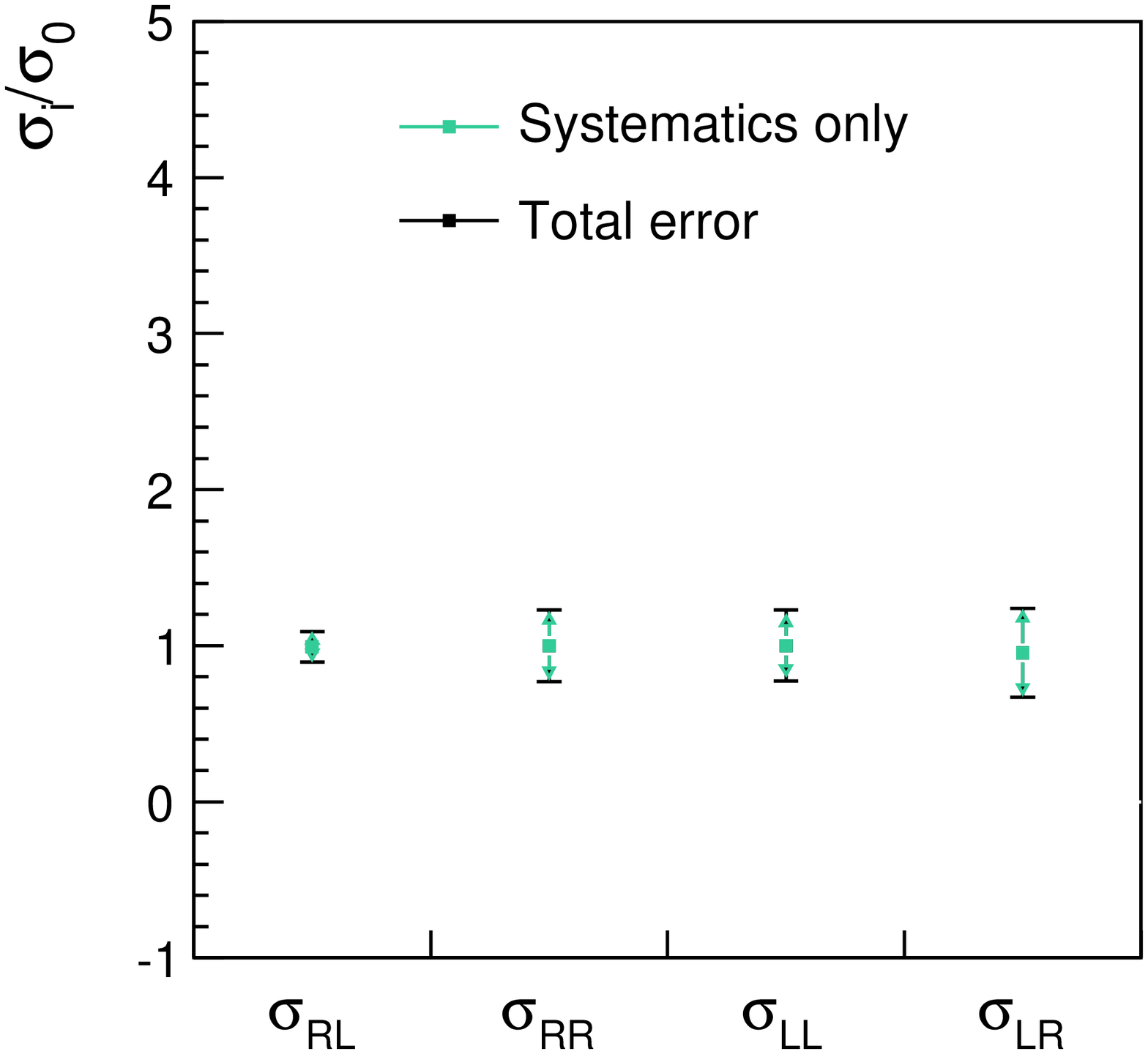,scale=.38}}
  \put(0.0, 7.0){\epsfig{file=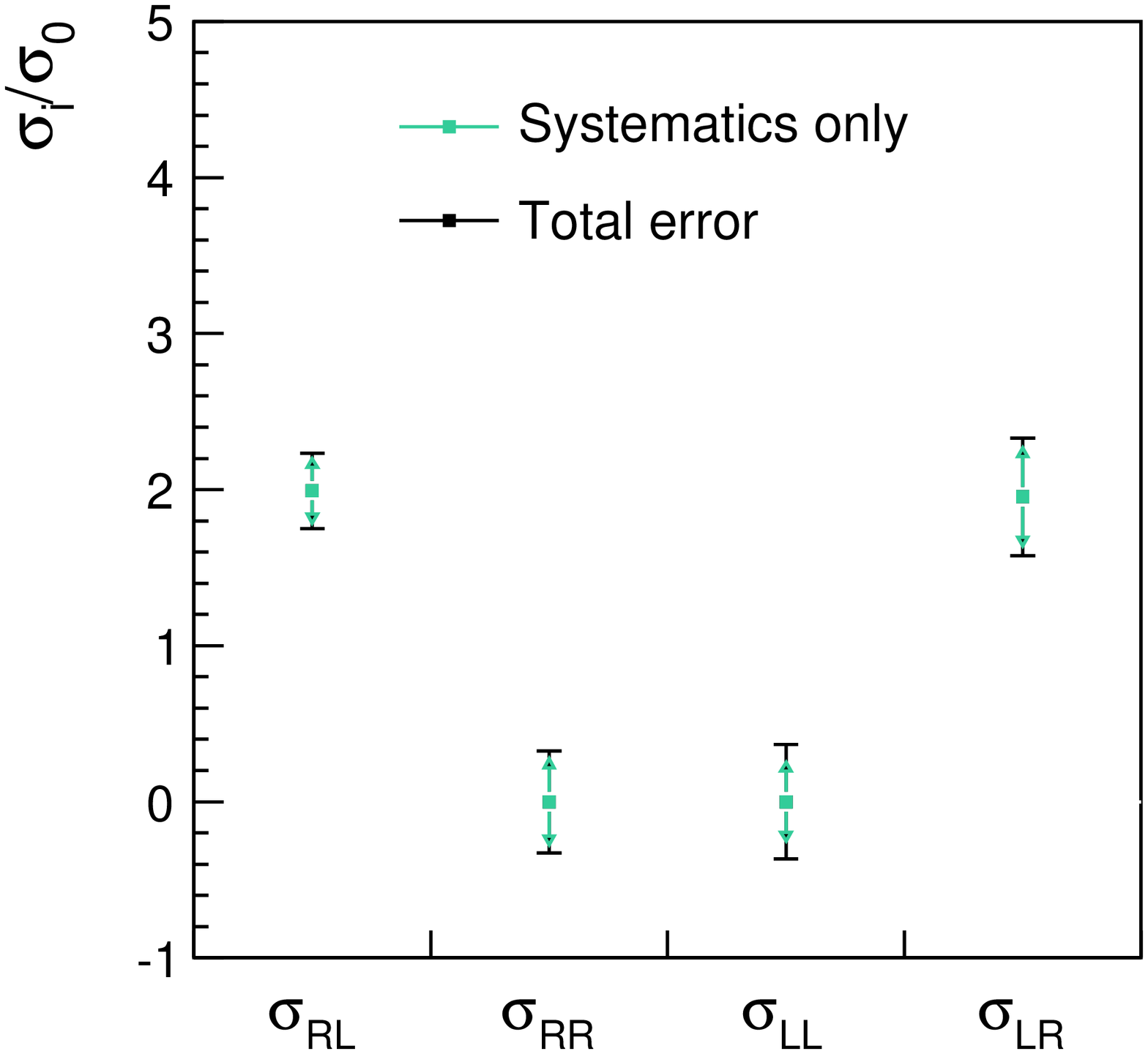,scale=.38}}
  \put(8.5, 7.0){\epsfig{file=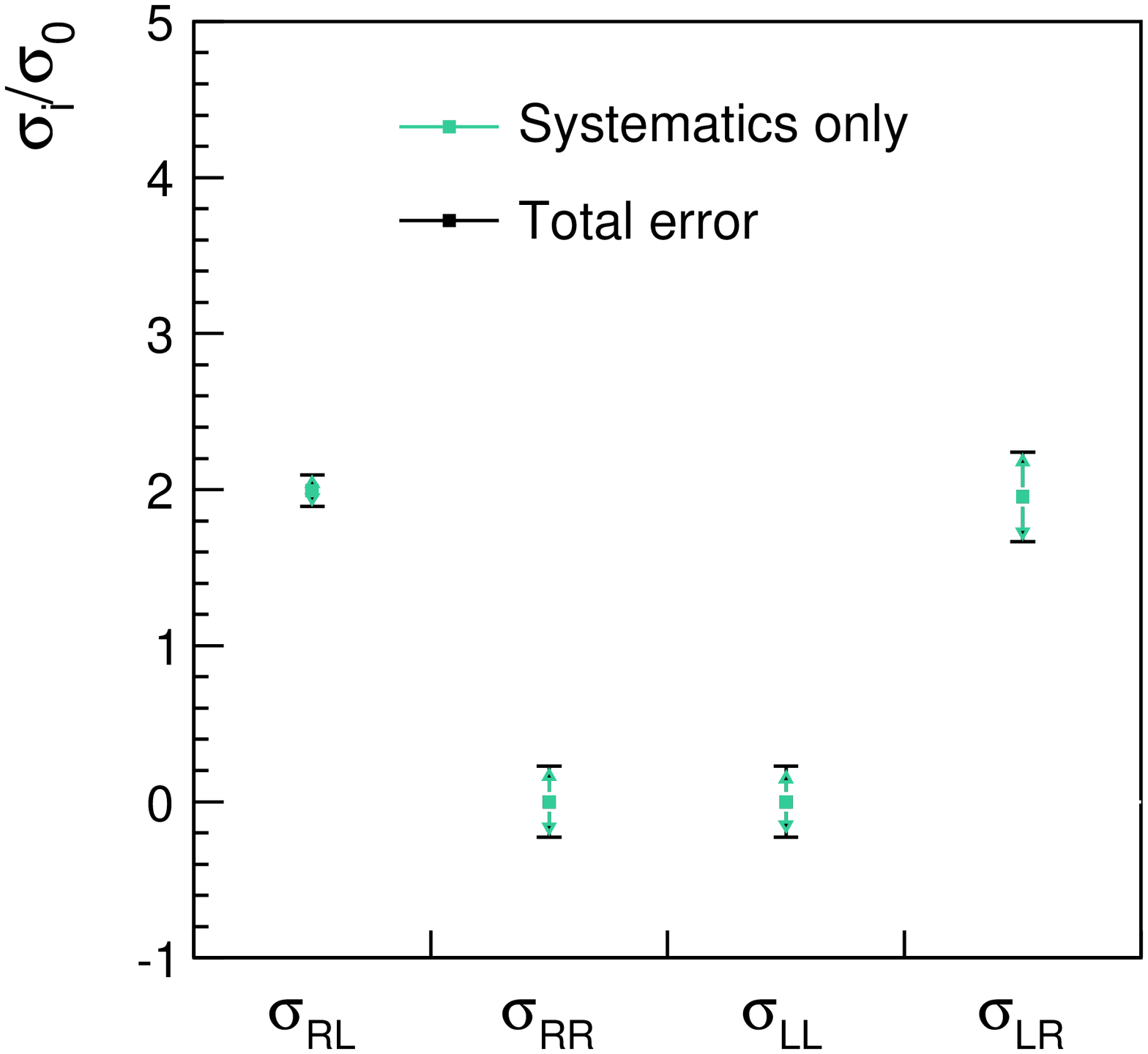,scale=.38}}
  \put(0.0, 0.0){\epsfig{file=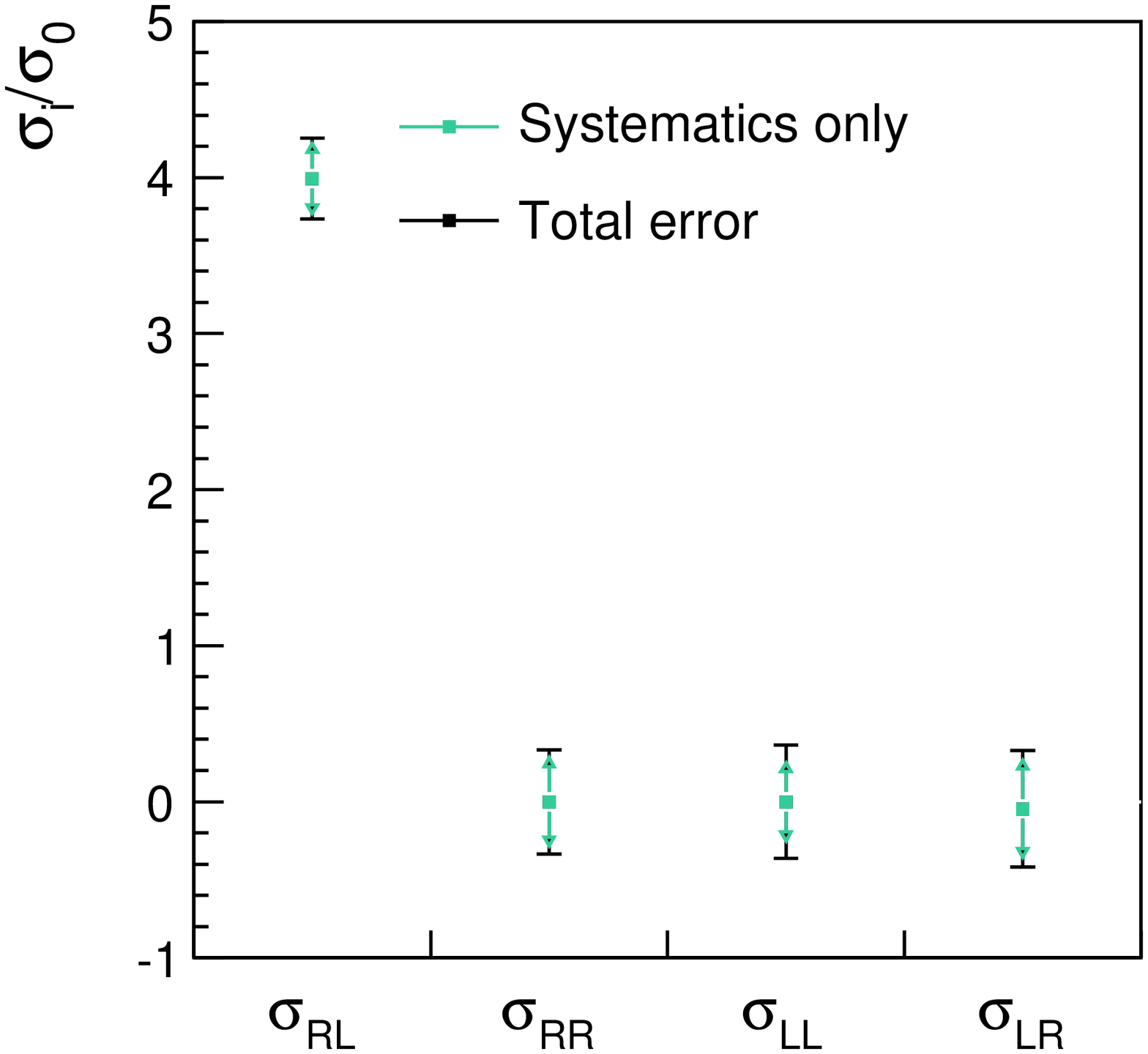,scale=.38}}
  \put(8.5, 0.0){\epsfig{file=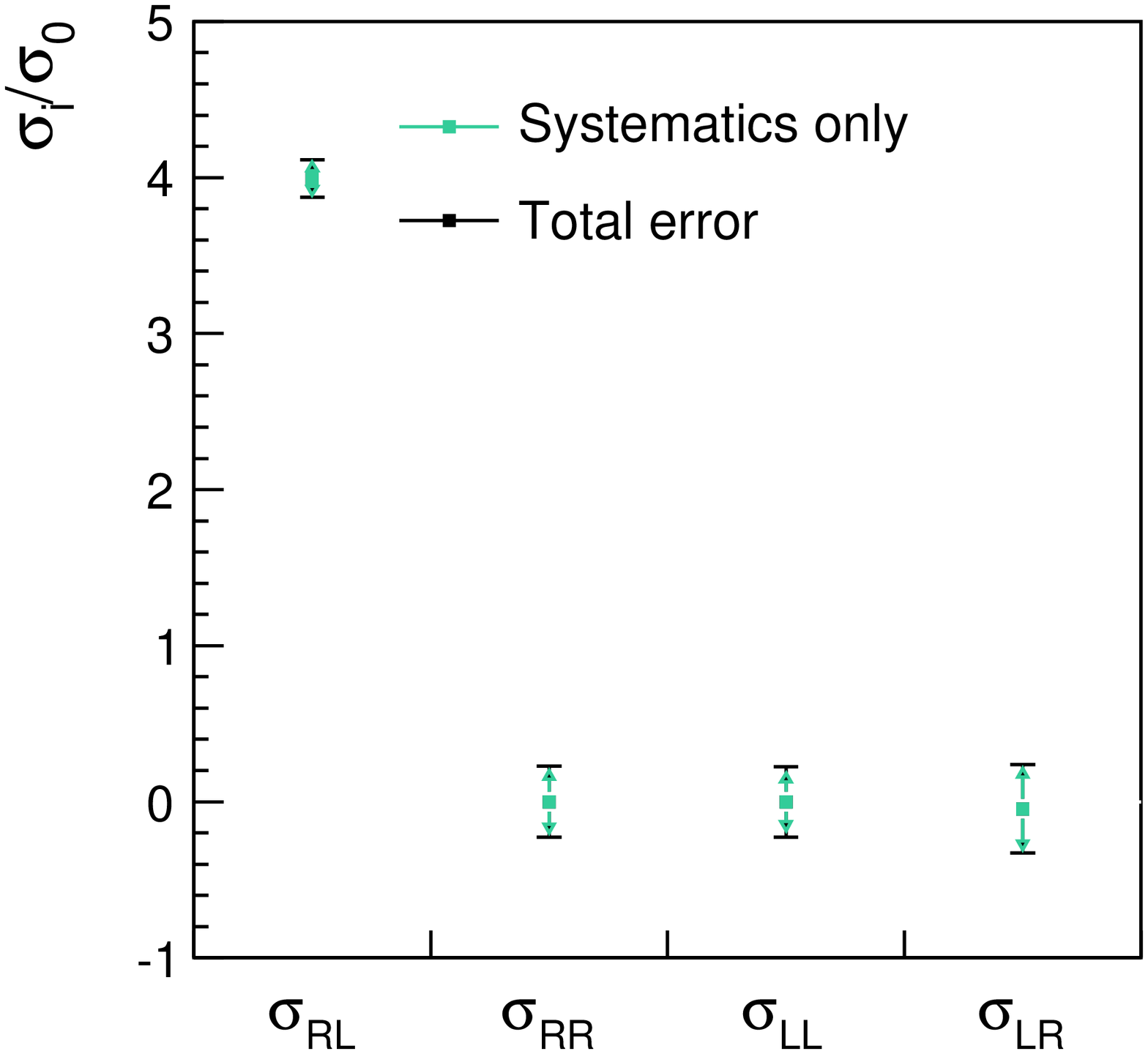,scale=.38}}   
  \put(0.0,15.0){(a)}   
  \put(8.5,15.0){(b)} 
  \put(0.0,8.0){(c)}   
  \put(8.5,8.0){(d)}  
  \put(0.0,1.0){(e)}   
  \put(8.5,1.0){(f)} 
  \end{picture}
  \caption{\label{fig:polxsect} Fully polarised cross-sections $\sigma_{\lbrace R,L\rbrace}$ measured 
    within three WIMP scenarios: (a)+(b) ``Equal'', (c)+(d) ``Helicity'', (e)+(f) ``Anti-SM'', 
    and for two different absolute values of the positron polarisation: (a)+(c)+(e) $|\Pp|=30\%$,
    (b)+(d)+(f) $|\Pp|=60\%$. 
    The cross-sections are normalised to the input unpolarised cross-section $\sigma_0$.}
\end{figure}

The results for the three studied coupling scenarios and a WIMP mass of $150$~GeV are listed in terms 
of the fully polarised cross-sections $\sigma_{\lbrace R,L\rbrace}$ in Table~\ref{Table:XSecPolMeas1} 
and shown in Figures~\ref{fig:polxsect}(a) to (f). Independently of the coupling scenario, the 
precision on the fully polarised cross-sections is dominated by the uncertainty on the beam 
polarisation as it enters the extrapolation to $|\Pe|=|\Pe|=100\%$. This can be seen from the fact that
the uncertainty reduces considerably when the positron polarisation is increased from $30\%$ to $60\%$
(right column in Table~\ref{Table:XSecPolMeas1}). About the same improvement is obtained by reducing 
the uncertainty of the polarisation measurement from $0.25\%$ to $0.1\%$ (values in parentheses in 
Table~\ref{Table:XSecPolMeas1}). In any of the cases, the coupling scenarios can be clearly separated from
each other.

\begin{table}[!h]
  \centering
  \renewcommand{\arraystretch}{1.10}
  \begin{tabular*}{\textwidth}{l@{\extracolsep{\fill}} p{6mm} rl p{4mm} rl}  
    \hline\hline 
    \multicolumn{7}{c}{\quad } \\[-4.8mm]
    Data scenario && \multicolumn{5}{c}{Unpolarised cross-section: $\;\sigma_{0} \;\pm\; {\rm stat} \;\pm\,{\rm sys}$\quad $(\pm\,{\rm total})$ [fb]} \\
    (simulated)   && \multicolumn{2}{c}{$(|\Pe|;\,|\Pp|)\;=\;(0.8;\,0.3)$} &&\multicolumn{2}{c}{$(|\Pe|;\,|\Pp|)\;=\;(0.8;\,0.6)$} \\[1pt]
    \hline\hline 
    \multicolumn{7}{c}{\quad } \\[-2mm]
    \multicolumn{7}{l}{ Assumed polarisation uncertainty\quad $\delta P/P = 0.25\%$ } \\ \hline
    \multicolumn{7}{c}{\quad } \\[-4.8mm]
    \equal &&    $99.0 \;\pm\; 2.8 \;\pm\; 4.3$  &  $(\pm\; 5.1)$   &&    $99.2 \;\pm\; 2.7 \;\pm\; 3.5$  &  $(\pm\; 4.4)$ \\
    \hel   &&    $99.1 \;\pm\; 2.3 \;\pm\; 4.0$  &  $(\pm\; 4.6)$   &&    $99.4 \;\pm\; 2.0 \;\pm\; 2.8$  &  $(\pm\; 3.4)$ \\
    \anti  &&    $99.8 \;\pm\; 1.4 \;\pm\; 2.8$  &  $(\pm\; 3.2)$   &&    $99.7 \;\pm\; 1.1 \;\pm\; 2.1$  &  $(\pm\; 2.4)$ \\
    \hline 
    \multicolumn{7}{c}{\quad } \\ 
    \multicolumn{7}{l}{ Assumed polarisation uncertainty\quad $\delta P/P = 0.10\%$ } \\ \hline
    \multicolumn{7}{c}{\quad } \\[-4.8mm]
    \equal &&    $99.0 \;\pm\; 2.6 \;\pm\; 2.0$  &  $(\pm\; 3.3)$   &&    $99.1 \;\pm\; 2.6 \;\pm\; 1.9$  &  $(\pm\; 3.2)$ \\
    \hel   &&    $99.1 \;\pm\; 2.3 \;\pm\; 2.0$  &  $(\pm\; 3.0)$   &&    $99.3 \;\pm\; 2.0 \;\pm\; 1.8$  &  $(\pm\; 2.6)$ \\
    \anti  &&    $99.6 \;\pm\; 1.4 \;\pm\; 1.8$  &  $(\pm\; 2.3)$   &&    $99.7 \;\pm\; 1.2 \;\pm\; 1.7$  &  $(\pm\; 2.1)$ \\
    \hline 
  \end{tabular*}
  \caption[Measurement results of unpolarised cross-section in three studied scenarios]{\it 
    Measured unpolarised cross-section $\sigma_{0}$ by a combination of cross-section measurements 
    with polarised beams for an integrated luminosity of $\lum =500\;\fb$.}
  \label{Table:XSecUnPolMeas}
\end{table}

The unpolarised cross-section $\sigma_0$ can be measured by combining all four measurements without the need
to extrapolate to $|\Pe|=|\Pe|=100\%$, which reduces impact of the polarimeter uncertainties 
significantly. Table~\ref{Table:XSecUnPolMeas} gives the achievable precisions on $\sigma_0$
in the three different coupling scenarios. Even though its impact is reduced, the polarisation
uncertainty still dominates the total error of $4$ to $5$~fb for $\delta P/P = 0.25\%$ , which 
can be considerably reduced to $2$ to $3$~fb when assuming $\delta P/P = 0.1\%$.

\begin{figure}[p]
  \begin{picture}(16.0, 22.0)
    \put( 0.00, 0.00)  {\epsfig{file=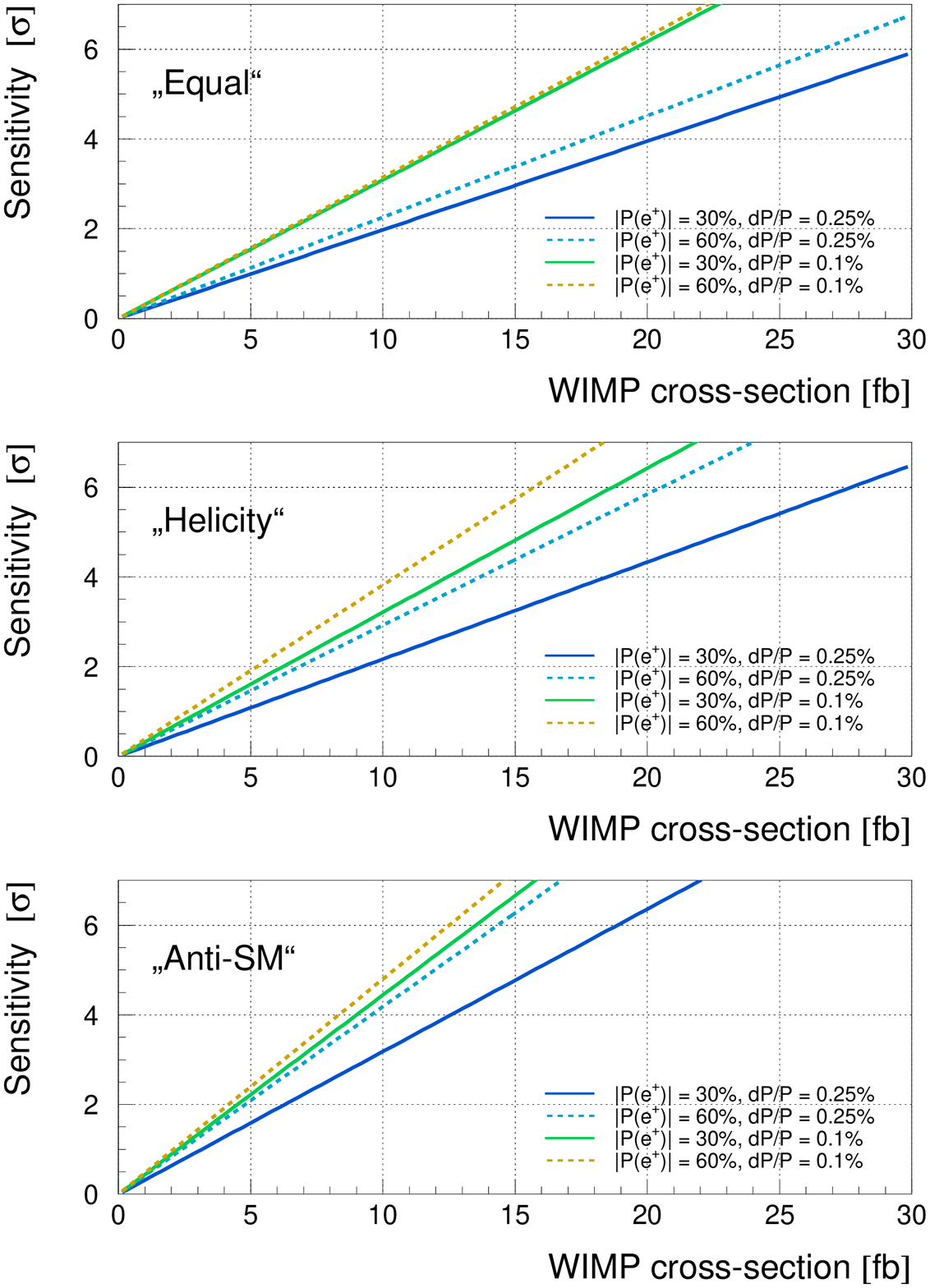, clip= , width=1.00\linewidth}}
  \end{picture}
  \caption{\label{fig:sens} \it Sensitivity in terms of standard deviations $\sigma$ as a function of the unpolarised WIMP cross-section for the three coupling scenarios and different assumptions on the positron polarisation and on the precision of the polarisation measurement.}
\end{figure}
These numbers have been derived based on an unpolarised cross-section of $\sigma_0 = 100$~fb. 
However the measurements are dominated by systematic uncertainties and by the statistical uncertainty on the background. Therefore the derived uncertainties are to a large extent independent of the actual 
signal cross-section. Figure~\ref{fig:sens} shows the sensitivity in terms of standard deviations ($\sigma$) as a function of the unpolarised WIMP cross-section for the three coupling scenarios and different assumptions on the positron polarisation and on the precision of the polarisation measurement. In absence of a signal in the data, cross-sections above $8.2$ ($3.6$)~fb can be excluded at $90\%$~CL in the worst (best) case. Cross-sections above $25$ ($11$)~fb could be observed at the $5\sigma$ level in the worst (best) case, which is in reasonable agreement with the minimal observable cross-section found in~\cite{bib:taikan} based on statistical uncertainties only.

\subsection{Mass Measurement and Extraction of the Dominant Partial Wave}
For the measurement of the WIMP mass and the extraction of the dominant partial wave, an integrated luminosity of $500$~fb$^{-1}$ 
of the $(|\Pe|;\,-|\Pp|)$ configuration is assumed\footnote{This is effectively an additional $300$~fb$^{-1}$ with respect to the data
set assumed for the cross-section measurement, which corresponds to a little more than one year of operation time of the ILC
after the initial four years.}.
The WIMP mass is determined by comparing template photon energy spectra for background and signals of different masses
against the ``data'' spectrum and by searching the mass for which the $\chi^{2}$ is minimised. This method 
is much more powerful than a mere determination of the maximal photon energy, because near the endpoint of the spectrum the signal is buried in the fluctuations of the huge Standard Model background, as can be seen in
Figure~\ref{fig:egammaSB}(b). 

\begin{figure}[htb]
  \begin{picture}(16.0, 8.0)
    \put( 0.00, 0.00)  {\epsfig{file=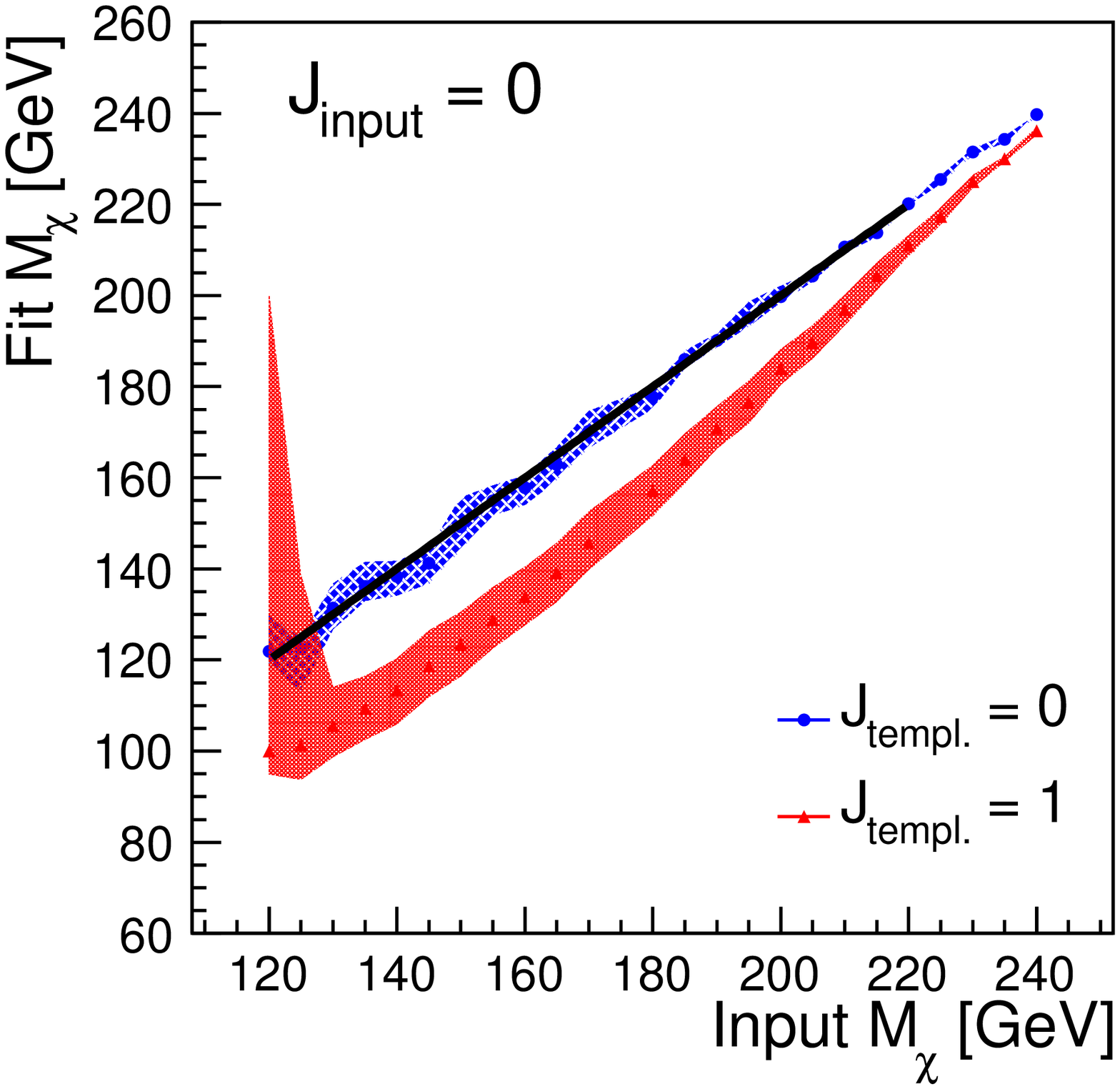, clip= , width=0.5\linewidth}}
    \put( 8.00, 0.00)  {\epsfig{file=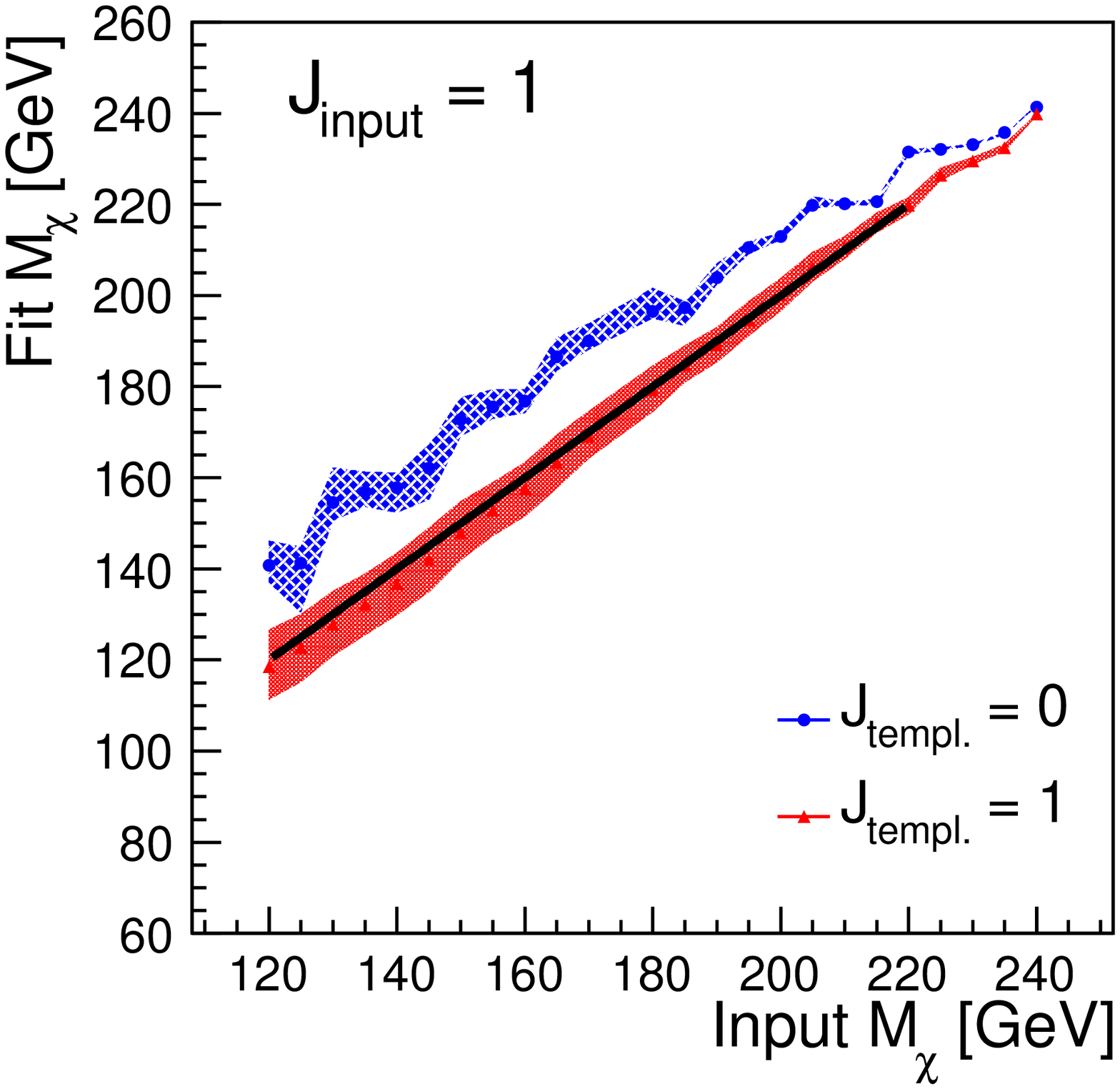, clip= , width=0.5\linewidth}}
    \put( 0.00, 0.00)  {(a)}
    \put( 8.00, 0.00)  {(b)}
  \end{picture}
  \caption{\label{fig:fitmass} \it Reconstructed vs true WIMP mass for the right and wrong partial wave assumptions in the templates.
  (a) true s-wave, (b) true p-wave.}
\end{figure}

The shape of the photon energy spectrum below the endpoint however also depends on the dominant partial wave
of the production, c.f. Figure~\ref{fig:dsigmadE}(b). Figure~\ref{fig:fitmass} shows the impact of the partial
wave assumption in the templates on the fitted WIMP mass. While the reconstructed mass follows the true mass
nicely when the correct partial wave templates are used, the wrong templates yield fitted masses either $20$~GeV 
too low ((a), true s-wave) or too high ((b), true p-wave). 

\begin{figure}[hp]
  \begin{picture}(16.0, 22.0)
  \put(0.0,14.0) {\epsfig{file=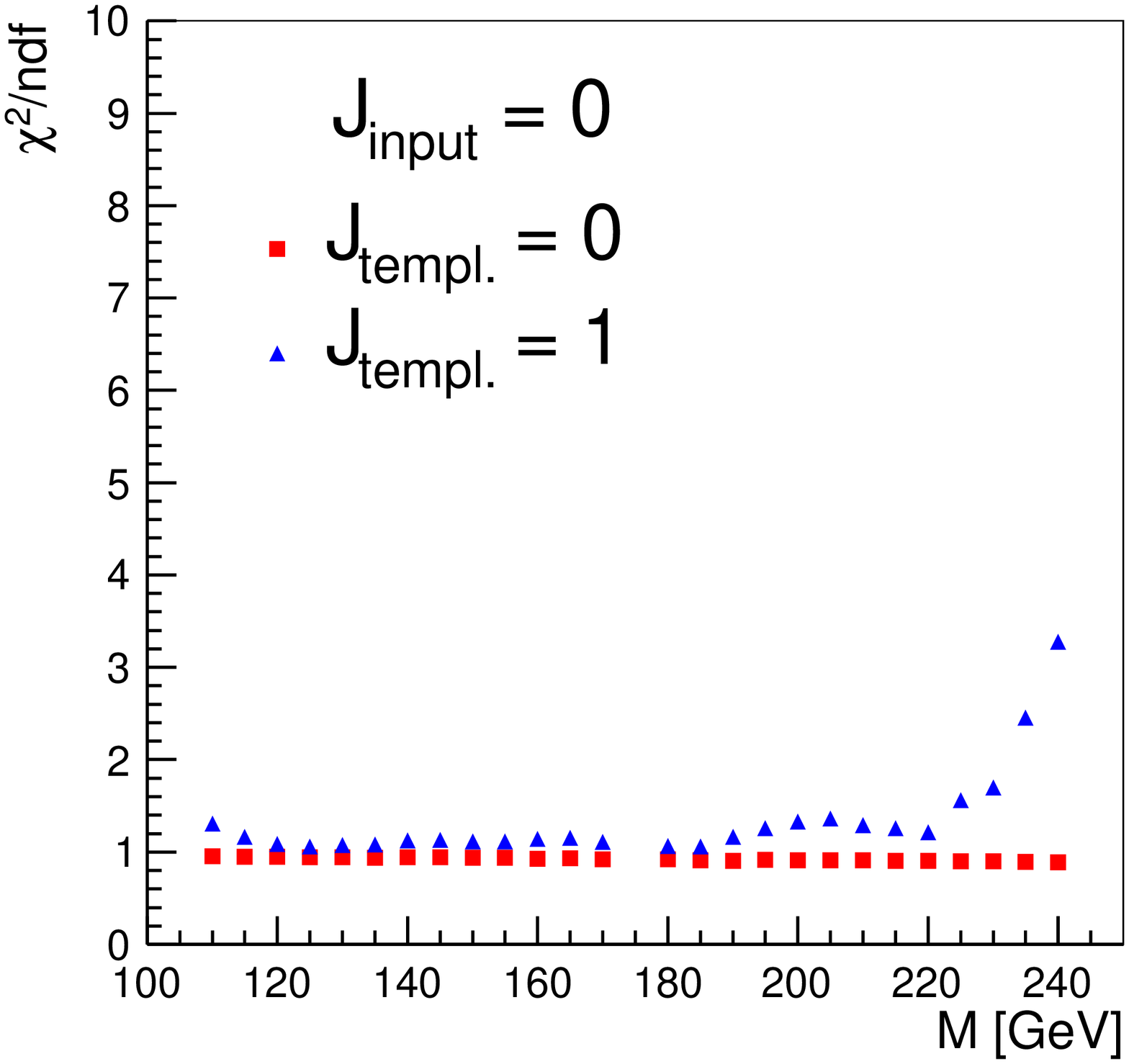,scale=.37}}
  \put(8.5,14.0) {\epsfig{file=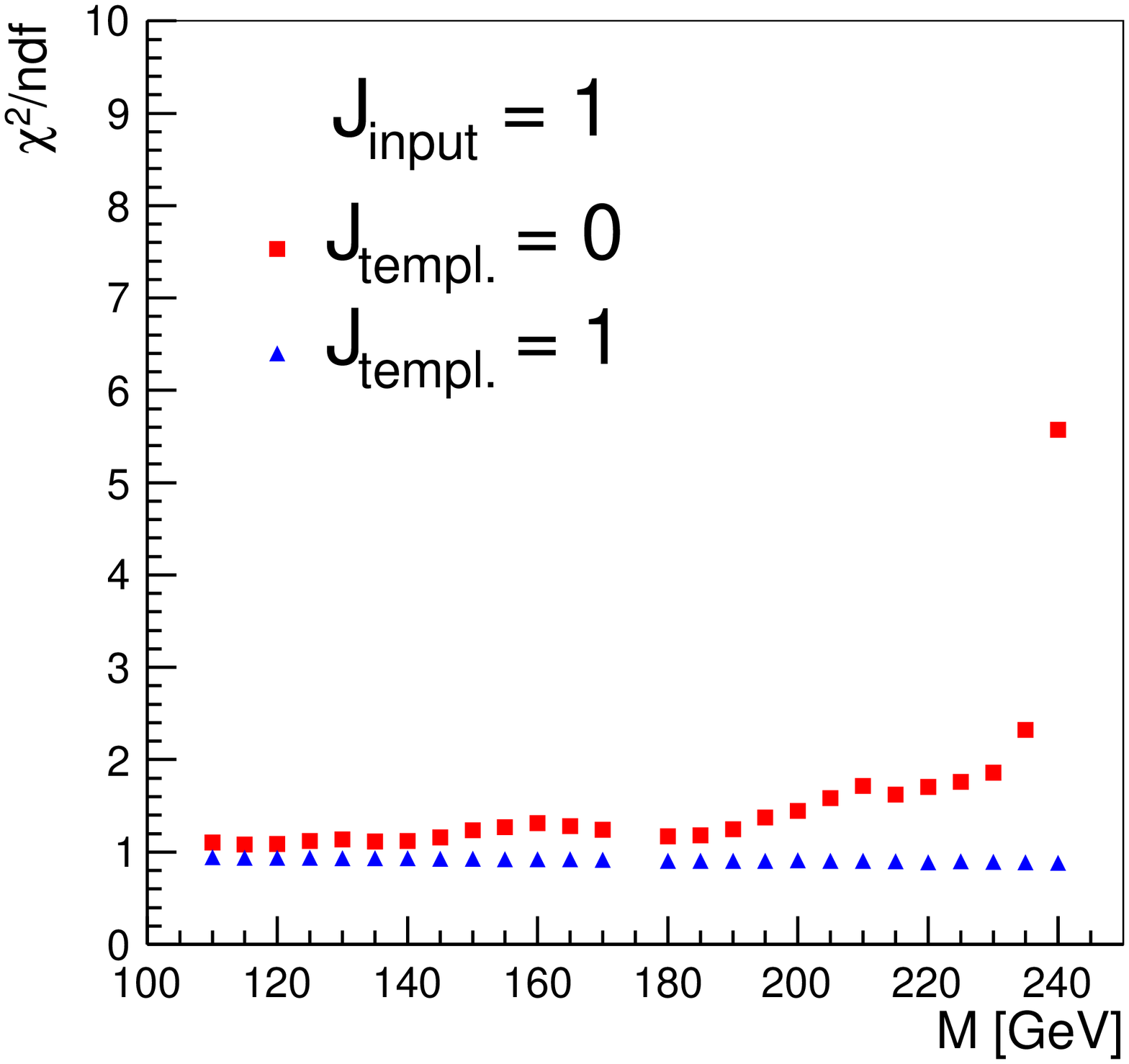,scale=.37}}
  \put(0.0, 7.0) {\epsfig{file=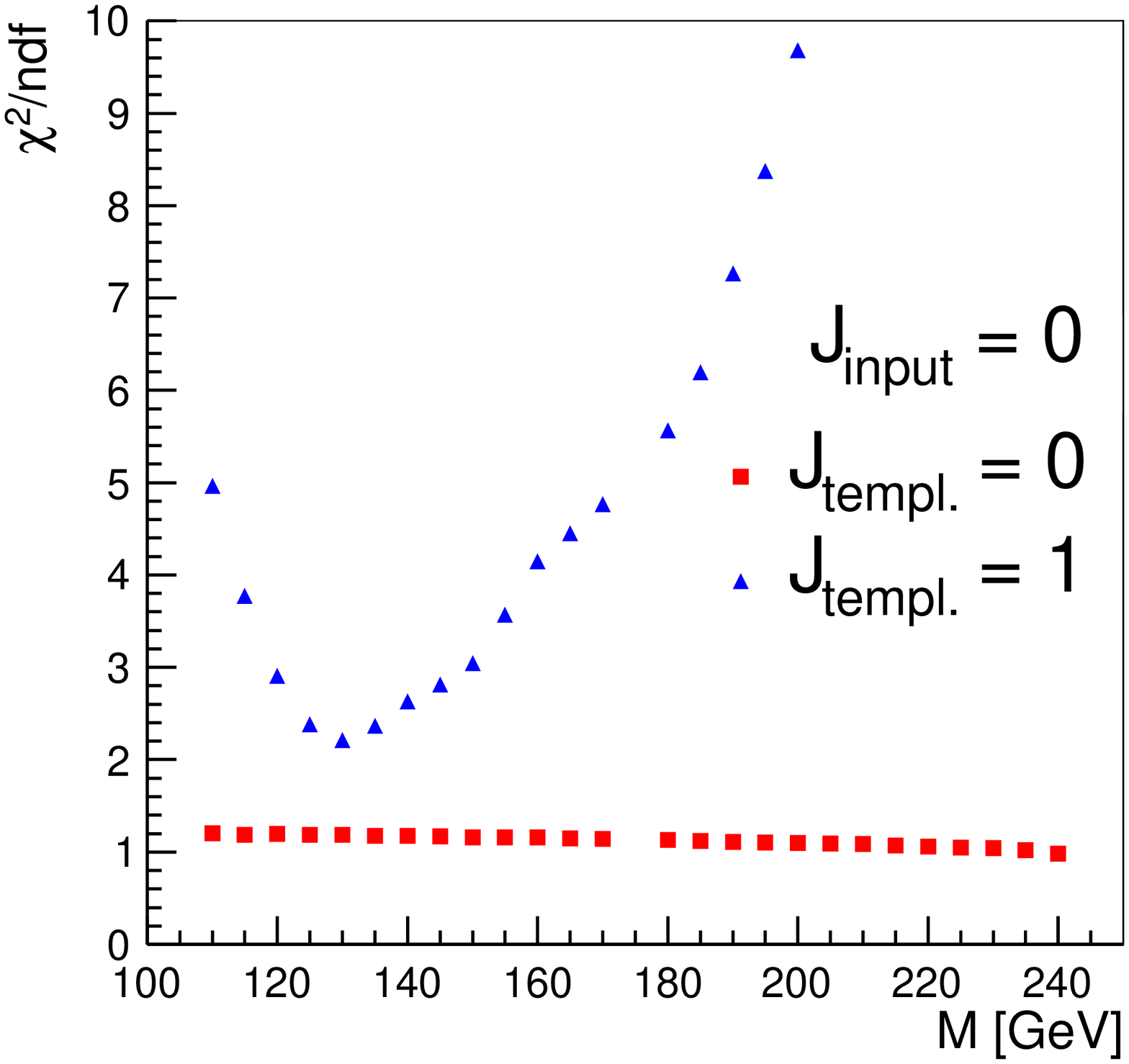,scale=.37}}
  \put(8.5, 7.0) {\epsfig{file=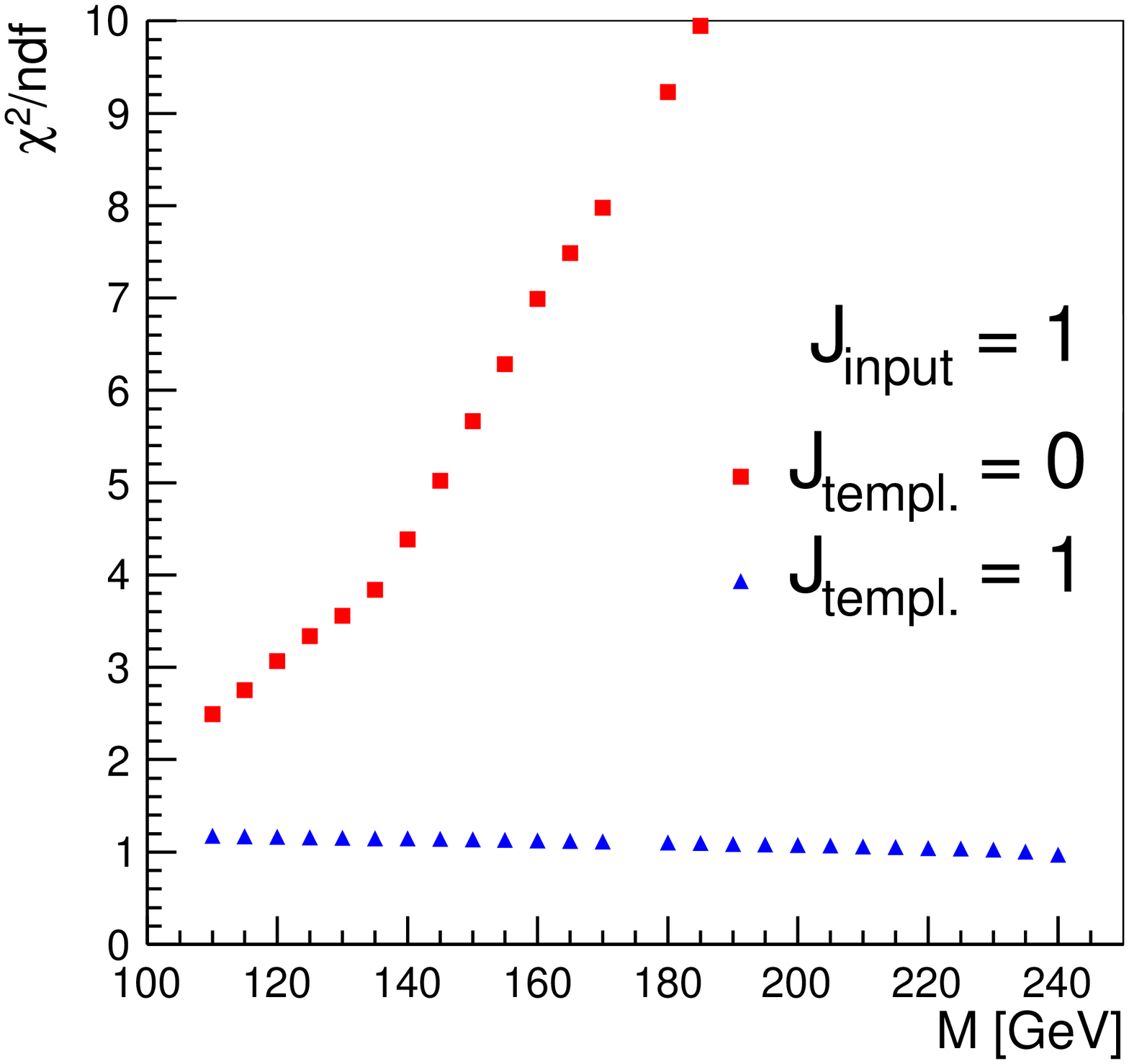,scale=.37}}
  \put(0.0, 0.0) {\epsfig{file=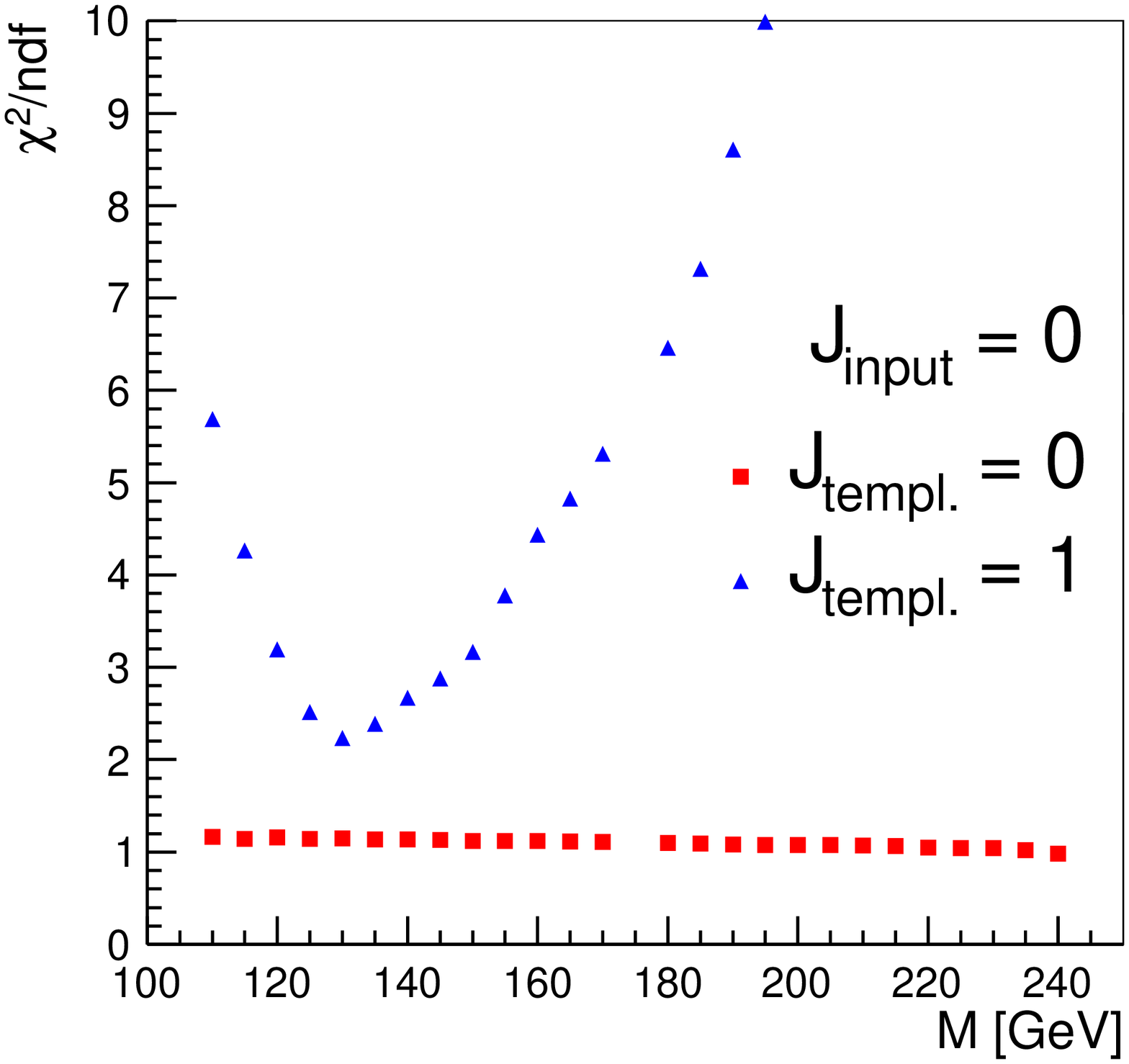,scale=.37}}
  \put(8.5, 0.0) {\epsfig{file=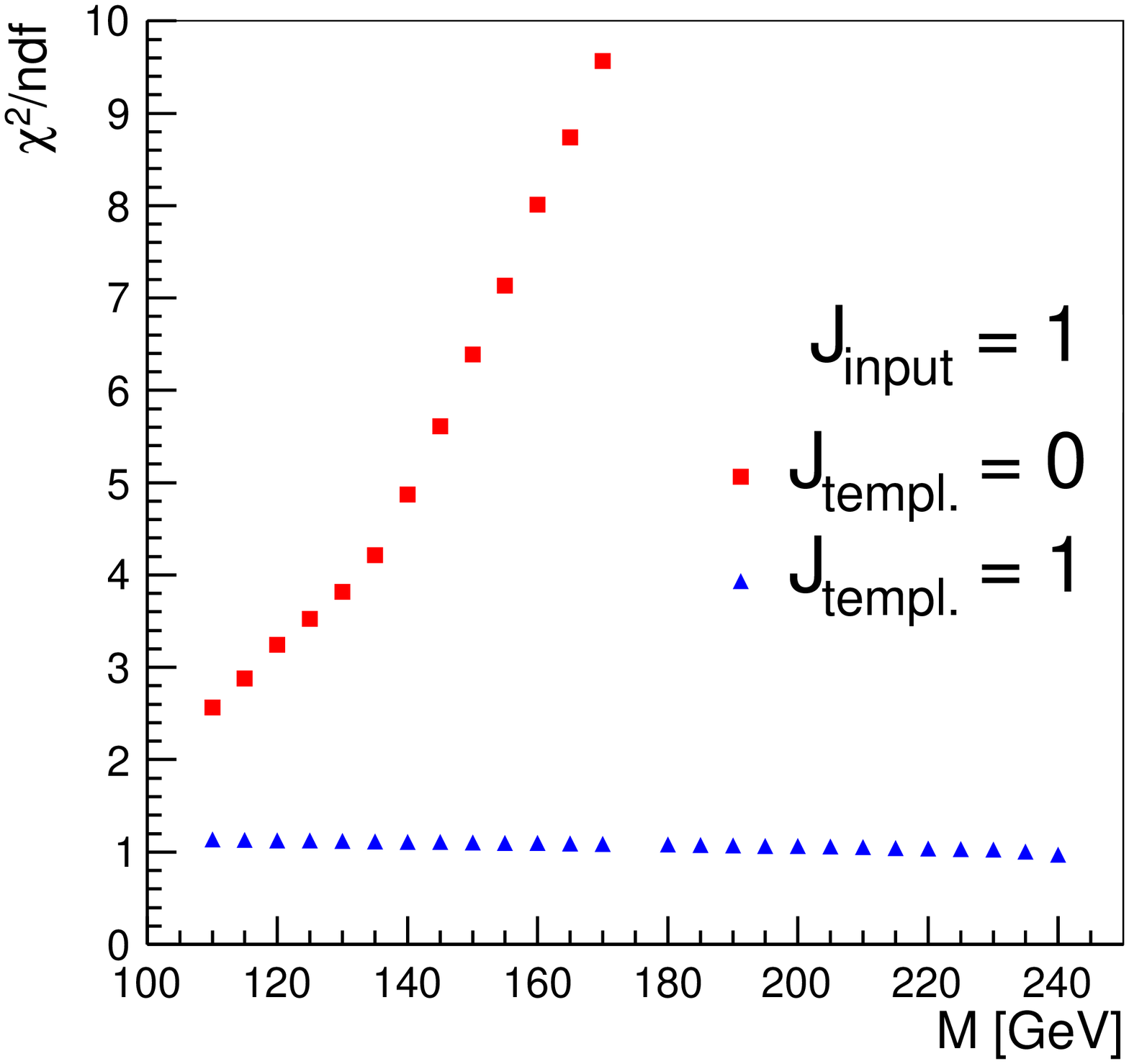,scale=.37}}   
  \put(0.0,14.0){(a)}   
  \put(8.5,14.0){(b)} 
  \put(0.0,7.0){(c)}   
  \put(8.5,7.0){(d)}  
  \put(0.0,0.0){(e)}   
  \put(8.5,0.0){(f)} 
  \end{picture}
  \caption{\label{fig:chi2mass} \it $\chi^2$/ndf of the template fit using the right and wrong partial wave assumptions as a function
  of the true WIMP mass. (a)+(b): $|\Pp| = 0$, (c)+(d): $|\Pp| = 30\%$, (e)+(f): $|\Pp| = 60\%$}
\end{figure}

However the correct assumption on the dominant partial wave can be easily determined over the full WIMP mass range by comparing the $\chi^2$ per degree of freedom
of the template fits if positron polarisation is available. This can be seen in Figure~\ref{fig:chi2mass},
where the $\chi^2$/ndf is displayed as a function of the true WIMP mass for the right and wrong partial wave hypothesis. In the upper row,
the positron polarisation is set to  $|\Pp| = 0$, and in this case no convincing difference in the $\chi^2$ values is obtained. For 
$|\Pp| = 30\%$ (middle row) and $|\Pp| = 60\%$ (bottom row), the wrong assumption yields a $\chi^2$/ndf significantly larger than unity.
Since for this test it is crucial to have a simulated sample corresponding to the actually assumed intergrated luminosity, the limited
amount of events available from full simulation has been augmented by a fast simulation sample obtained from an ILD version of 
{\sc SGV}~\cite{bib:sgv} for this (and only this) figure.

\begin{table}[!h]
  \centering
  \renewcommand{\arraystretch}{1.10}
  \begin{tabular*}{\textwidth}{l@{\extracolsep{\fill}} rr}  
    \hline\hline 
    \multicolumn{3}{c}{\quad } \\[-4.8mm]
    Mass   & \multicolumn{2}{c}{WIMP mass: $\;\pm\,$stat.  $\pm\,\delta E\,$(scale)  $\pm\,\delta E\,$(shape)$\;$ (total) [GeV]} \\
    \protect[GeV\protect]\hspace*{5mm}
    $\,$   &  $(\Pe;\,\Pp)\,=\,(0.8;\,-0.3)$    & $(\Pe;\,\Pp)\,=\,(0.8;\,-0.6)$ \\[1pt]
    \hline\hline 
    \multicolumn{3}{c}{\quad} \\[-2.5mm]
    \multicolumn{3}{l}{\quad\equal\quad scenario } \\ \hline
    \multicolumn{3}{c}{\quad} \\[-4.8mm]
    $120$  &  $2.48 \pm0.07 \pm1.90\;\; (3.12)$  &  $2.24 \pm0.07 \pm1.90\;\; (2.93)$ \\
    $150$  &  $1.98 \pm0.05 \pm1.46\;\; (2.46)$  &  $1.83 \pm0.05 \pm1.45\;\; (2.33)$ \\
    $180$  &  $1.69 \pm0.03 \pm1.00\;\; (1.96)$  &  $1.57 \pm0.03 \pm1.00\;\; (1.86)$ \\
    $210$  &  $0.79 \pm0.02 \pm0.54\;\; (0.96)$  &  $0.74 \pm0.02 \pm0.54\;\; (0.91)$ \\
    \hline 
    \multicolumn{3}{c}{\quad} \\[-2.5mm]
    \multicolumn{3}{l}{\quad\hel\quad scenario } \\ \hline
    \multicolumn{3}{c}{\quad} \\[-4.8mm]
    $120$  &    $1.92 \pm0.07 \pm1.89\;\; (2.70)$  &  $1.53 \pm0.07 \pm1.89\;\; (2.43)$ \\
    $150$  &    $1.62 \pm0.05 \pm1.46\;\; (2.18)$  &  $1.23 \pm0.05 \pm1.45\;\; (1.90)$ \\
    $180$  &    $1.36 \pm0.03 \pm1.00\;\; (1.69)$  &  $0.94 \pm0.03 \pm1.00\;\; (1.37)$ \\
    $210$  &    $0.67 \pm0.02 \pm0.54\;\; (0.87)$  &  $0.59 \pm0.02 \pm0.54\;\; (0.80)$ \\
    \hline 
    \multicolumn{3}{c}{\quad} \\[-2.5mm]
    \multicolumn{3}{l}{\quad\anti\quad scenario } \\ \hline
    \multicolumn{3}{c}{\quad} \\[-4.8mm]
    $120$  &    $1.04 \pm0.07 \pm1.88\;\; (2.15)$  &  $0.82 \pm0.07 \pm1.88\;\; (2.05)$ \\
    $150$  &    $0.81 \pm0.05 \pm1.45\;\; (1.66)$  &  $0.72 \pm0.05 \pm1.44\;\; (1.61)$ \\
    $180$  &    $0.66 \pm0.03 \pm1.00\;\; (1.19)$  &  $0.37 \pm0.03 \pm1.00\;\; (1.06)$ \\
    $210$  &    $0.16 \pm0.02 \pm0.55\;\; (0.59)$  &  $0.09 \pm0.02 \pm0.55\;\; (0.59)$ \\
    \hline 
  \end{tabular*}
  \caption{\it 
    Statistical and systematic uncertainties on the measured WIMP masses 
    for an integrated luminosity of $\lum = 500\;\fb$ 
    in the three coupling scenarios \equal, \hel{} and \anti{} 
    for three different polarisation configurations.}
  \label{tab:MassMeasuremnt}
\end{table}

\begin{figure}[hp]
  \begin{picture}(15.0, 21.0)
 \put(0.0,14.0){\epsfig{file=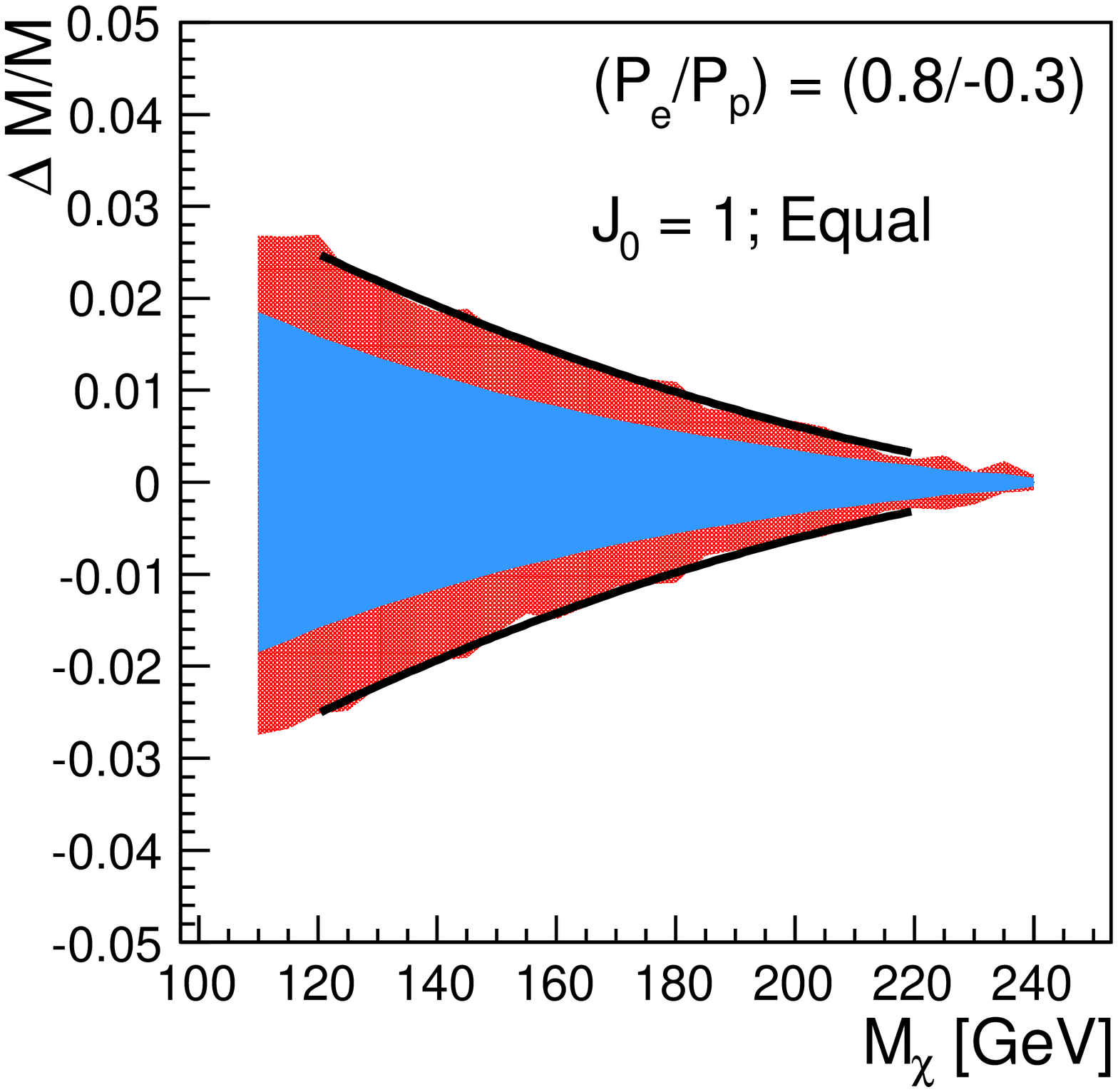, scale=.37}}
 \put(8.5,14.0){\epsfig{file=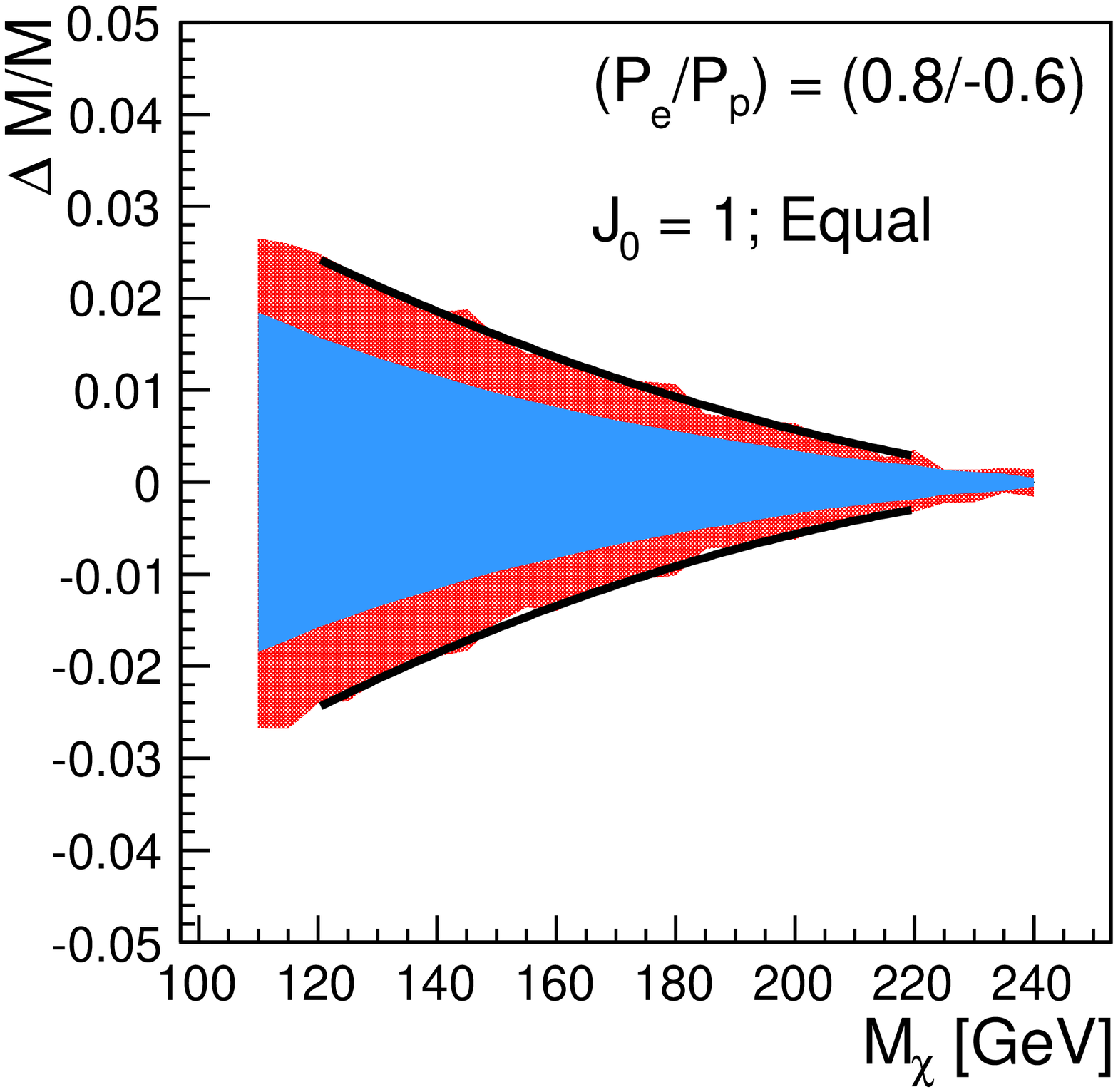, scale=.37}}  
 \put(0.0, 7.0){\epsfig{file=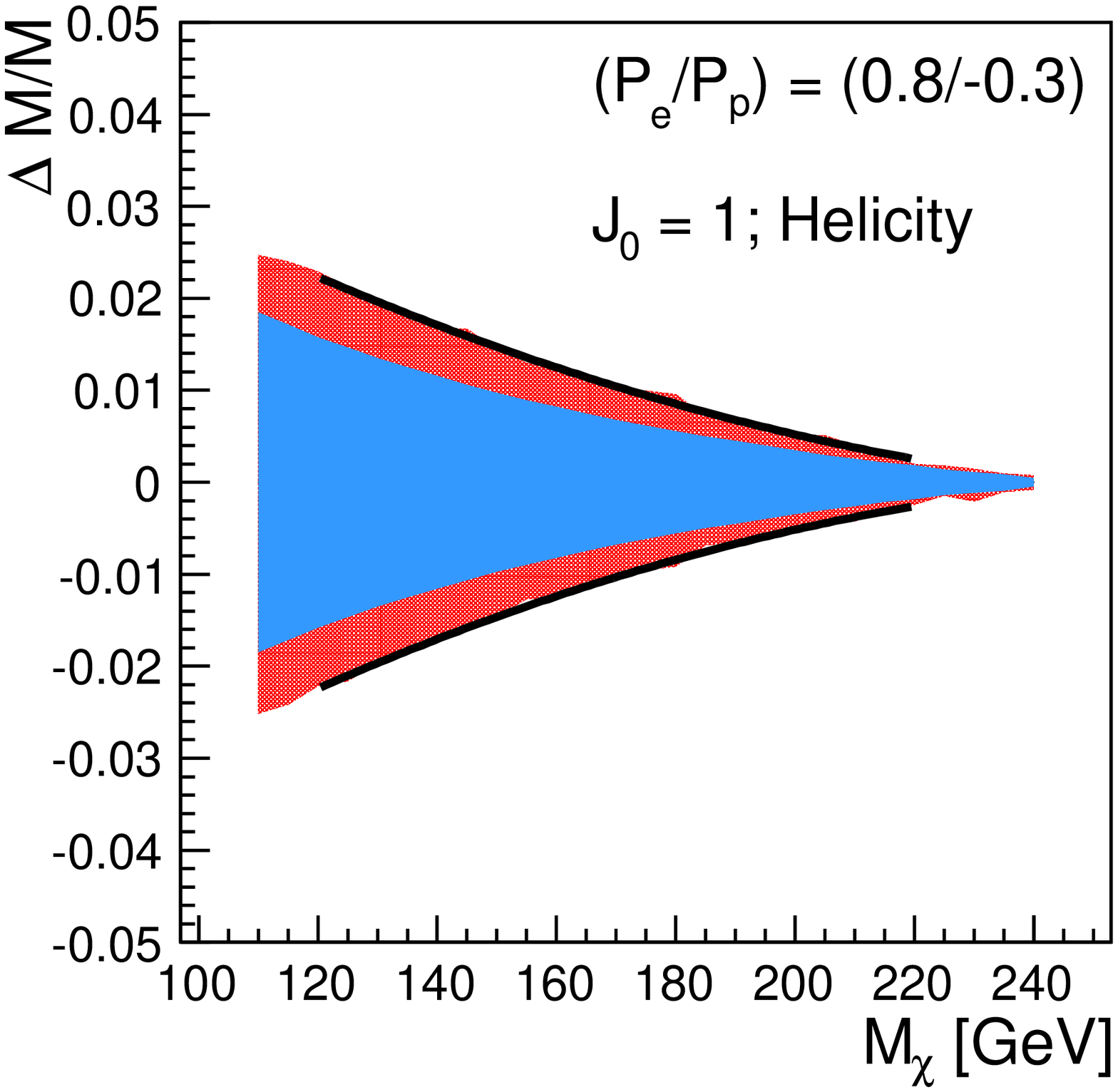, scale=.37}}
 \put(8.5, 7.0){\epsfig{file=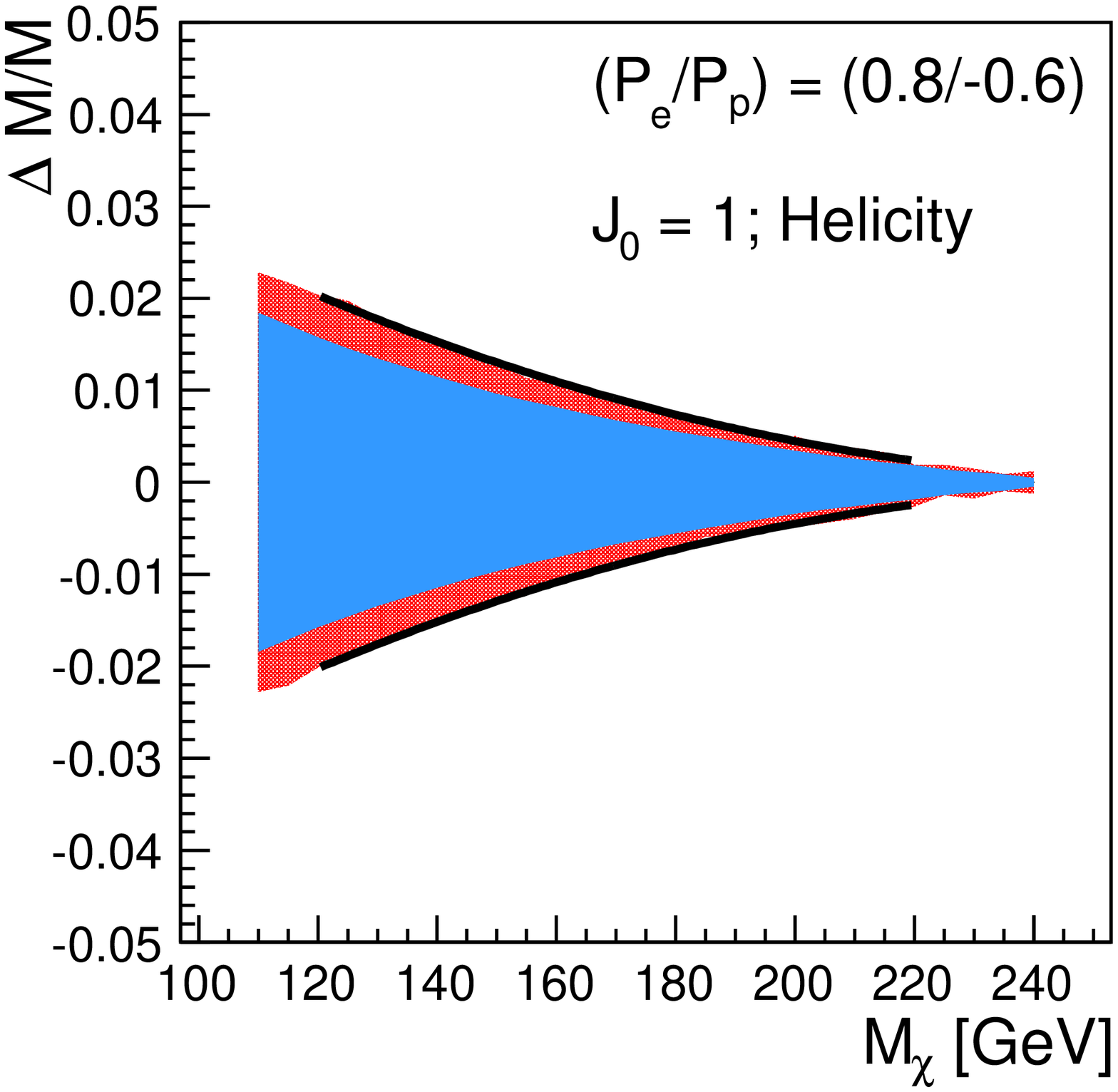, scale=.37}}  
 \put(0.0, 0.0){\epsfig{file=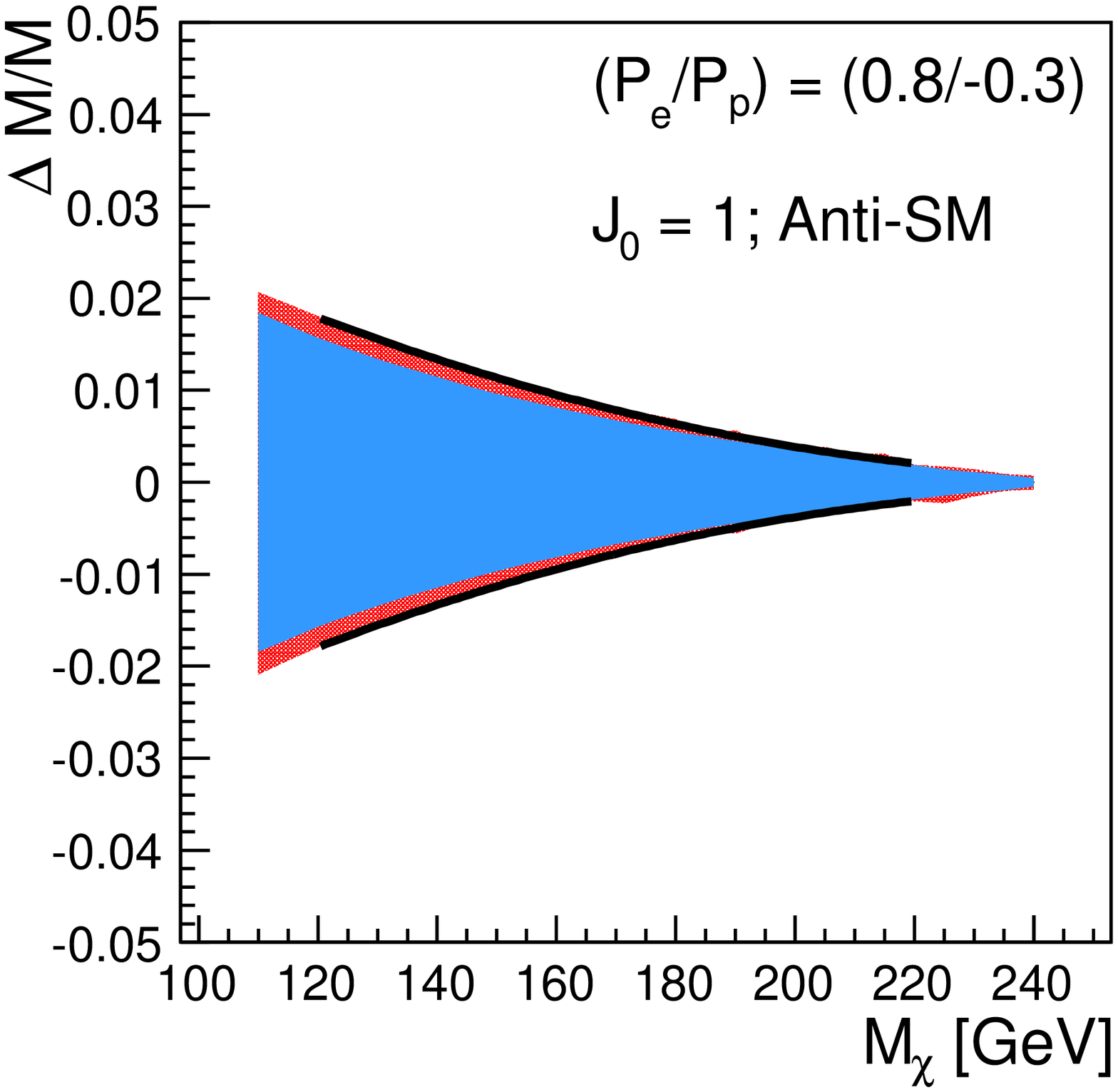, scale=.37}}
 \put(8.5, 0.0){\epsfig{file=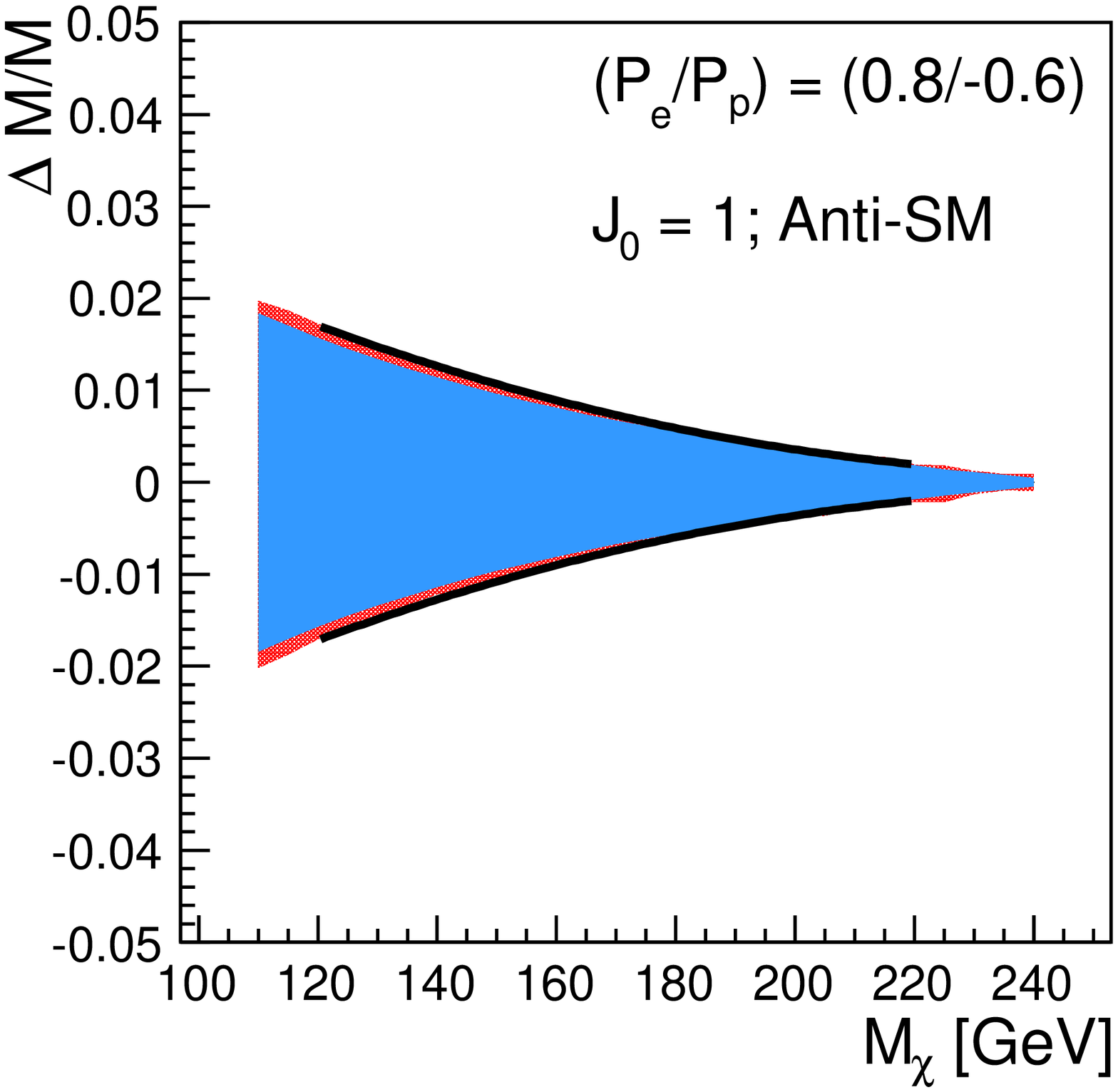, scale=.37}}  
 \put(0.0,14.0){(a)}
 \put(8.5,14.0){(b)}
 \put(0.0,7.0){(c)}
 \put(8.5,7.0){(d)}
 \put(0.0,0.0){(e)}
 \put(8.5,0.0){(f)}
  \end{picture}
  \caption{\label{fig:deltamass} \it Relative uncertainty on the reconstructed WIMP mass as a function of the true WIMP mass for the three different coupling scenarios and two different values of the positron polarisation. The blue area shows the systematic uncertainty and the red bands the additional statistical contribution.}
\end{figure}

After resolving the ambiguity due to the dominant partial wave, the main systematic uncertainty in the mass measurement is the shape
of the beam energy spectrum followed by the uncertainty on the beam energy scale. The former has been estimated from a comparison to fits with templates based on the beam energy spectrum obtained with the SB2009 beam parameter set. Conservatively, the full resulting shift in measured mass is included in the systematic errors quoted here. Thus 
the uncertainty of the mass measurement can be reduced by determining the beam parameters, for instance from the shape of the energy depositions of the $e^+e^-$ pair background in the forward detectors~\cite{bib:beampars}. This is forseen in the design of the ILC detectors, but has not been studied in this analysis. Table~\ref{tab:MassMeasuremnt} and Figure~\ref{fig:deltamass} display the statistical and systematic accuracies reached under these assumptions for the three coupling scenarios and two values of the positron polarisation. The total uncertainty ranges between $0.5$ and $3\%$. For higher WIMP masses the measurement becomes in general more  precise, because the photon energy spectrum is restricted to lower values and thus the signal is more prominent locally.

\section{Conclusions}  \label{sec:conclusion}
The photon plus missing energy signature at the International Linear Collider has been investigated in the context of a model-independent
characterisation of WIMPs. In view of the substantial Standard Model backgrounds to this signature, the analysis has been performed in
full detector simulation and with realistic assumptions on the beam properties and resulting systematic uncertainties. 
The unpolarised cross-section can be measured with an accuracy between $2$ and $5$~fb depending on the beam polarisation and the WIMP scenario. The helicity structure of the WIMP-fermion interaction can be obtained from the polarised cross-section. These measurements are systematically limited by the knowledge of the beam polarisation.

Via the shape of the photon energy spectrum, p- and s-wave production can be clearly distinguished if positron polarisation is available. Finally the mass of the WIMP can be determined with an accuracy between $0.5\%$ and $3\%$, depending on the WIMP parameters. These numbers are limited by the knowledge of the shape of the beam energy spectrum.

Beyond the model-independent WIMP approach, these results can also be used in any specific scenario beyond the Standard Model with invisible or nearly invisible massive particles in the kinematic reach of the ILC.

\section*{Acknowledgements}  \label{sec:acknowledge}
We thank Olaf Kittel and Ulrich Langenfeld helpful discussions and for 
providing the {\sc Fortran} implementation of the double differential 
tree--level cross--section for the process $e^+e^- \ra \nng$ 
in the Standard Model as a function of the center-of-mass energy and 
the beam polarisations. We are grateful to Katarzyna Wichmann and Anthony Hartin who provided the information on the pair background as well as to the complete ILCSoft and ILD Monte-Carlo production teams.

This work was supported by the German Science Foundation (DFG) via Emmy-Noether grant
Li~1560/1-1 and partially supported within the Collaborative Research Center 676 ``Particles, Strings and the Early Universe''.

\begin{footnotesize}


\end{footnotesize}


\end{document}